\documentclass[reprint,
groupedaddress,
nobibnotes,
 amsmath,amssymb,
aip,
]{revtex4-1}
\usepackage{epsfig}
\usepackage{graphicx}
\usepackage{dcolumn}
\usepackage{bm}
\usepackage{amsthm}
\usepackage{amsfonts}
\usepackage{color} 
\usepackage[usenames,dvipsnames]{xcolor}
\usepackage{amscd}
\usepackage{amsmath}
\usepackage{enumerate}
\usepackage{bbm}
\usepackage{epstopdf}
\usepackage[colorlinks=true,citecolor=blue,urlcolor=black]{hyperref}

\newcommand{\ie}{\textit{i.e.}}
\newcommand{\eg}{\textit{e.g.}}

\newcommand{\sket}[1]{{\ensuremath{\lvert#1\rangle}}}
\newcommand{\lket}[1]{{\ensuremath{\left\lvert#1\right\rangle}}}
\newcommand{\ket}[1]{\if@display\lket{#1}\else\sket{#1}\fi}

\newcommand{\sbra}[1]{{\ensuremath{\langle#1\rvert}}}
\newcommand{\lbra}[1]{{\ensuremath{\left\langle#1\right\rvert}}}
\newcommand{\bra}[1]{\if@display\lbra{#1}\else\sbra{#1}\fi}

\newcommand{\sbraket}[2]{{\ensuremath{\langle#1\rvert#2\rangle}}}
\newcommand{\lbraket}[2]{{\ensuremath{\left\langle#1\!\left\rvert\vphantom{#1}#2\right.\!\right\rangle}}}
\newcommand{\braket}[2]{\if@display\lbraket{#1}{#2}\else\sbraket{#1}{#2}\fi}

\newcommand{\sketbra}[2]{{\ensuremath{\lvert #1\rangle\!\langle #2\rvert}}}
\newcommand{\lketbra}[2]{{\ensuremath{\left\lvert #1\right\rangle\!\!\left\langle #2\right\rvert}}}
\newcommand{\ketbra}[2]{\if@display\lketbra{#1}{#2}\else\sketbra{#1}{#2}\fi}

\begin{document}
\title{Long distance measurement-device-independent quantum key distribution with entangled photon sources}

\author{Feihu Xu}
  \email{feihu.xu@utoronto.ca}

\author{Bing Qi}

\author{Zhongfa Liao}

\author{Hoi-Kwong Lo}

\affiliation{%
Centre for Quantum Information and Quantum Control,\\
Department of Physics and Department of Electrical \& Computer Engineering,\\
University of Toronto, Toronto,  Ontario, M5S 3G4, Canada, }

\date{\today}

\begin{abstract}
We present a feasible method that can make quantum key distribution (QKD) both ultra-long-distance and immune to all attacks in the detection system. This method is called measurement-device-independent QKD (MDI-QKD) \emph{with} entangled photon sources in the middle. By proposing a model and simulating a QKD experiment, we find that MDI-QKD \emph{with} one entangled photon source can tolerate 77dB loss (367km standard fiber) in the asymptotic limit and 60dB loss (286km standard fiber) in the finite-key case with state-of-the-art detectors. Our general model can also be applied to other non-QKD experiments involving entanglement and Bell state measurements.
\end{abstract}

\maketitle
The global quantum internet~\cite{kimble2008quantum} is believed to be the next-generation information processing platform promising an exponentially speed-up computation~\cite{kok2007linear} and a secure means of communication. The long-distance distribution of quantum states is a key ingredient for such a global platform, and recently, it has attracted significant scientific attention~\cite{yin2012quantum, ma2012quantum}. Among the applications of global quantum internet, quantum key distribution (QKD)~\cite{bennett1984quantum,ekert1991quantum} has been identified as the first technology in quantum information science to reach practical applications. Tremendous effort has been dedicated to creating a global QKD network during the past decade~\cite{QC:Gisin:2002,peev2009secoqc,sasaki2011field}. Nonetheless, a real-life QKD network is still limited by two important factors -- performance and security.

For performance, long-distance QKD remains challenging. In experiment, the maximal transmission distances are 200km through standard telecom fiber for the decoy-state BB84 protocol~\cite{dixon2008gigahertz,liu2010decoy} and 144km through free space for the entanglement based QKD~\cite{Ursin:144QKD}. In theory, the decoy-state BB84 protocol with excellent detectors can tolerate a maximal loss of around 50 dB in the asymptotic limit of an infinitely long key~\cite{rosenberg2009practical}; in practice, however, the finite-key effect~\cite{renner2005security} of the data transmission will substantially lower the tolerable loss, \eg, to less than 35 dB~\cite{cai2009finite}. On the other hand, the entanglement based QKD with sophisticated post-processing can in principle tolerate higher losses of 70 dB in the asymptotic limit and around 50 dB in the case of a finite key~\cite{ma2007quantum} (also, a more rigorous finite-key analysis will further decrease the tolerable loss to less than 40 dB~\cite{cai2009finite}). Nevertheless, we remark that all the above schemes are vulnerable to various detector side-channel attacks (see discussion below).

For security, the unconditional security of QKD has been rigorously proved based on the laws of quantum mechanics~\cite{Mayers:2001, Lo:1999, Shor:2000, QKD:Scarani:2009}. However, real-life implementations of QKD may contain overlooked imperfections, which are missing in the theoretical model of security proofs. By exploiting these imperfections, especially those in detectors, researchers have demonstrated various quantum attacks including time-shift attack~\cite{Yi:timeshift:2008}, phase-remapping attack~\cite{Xu:phaseremapping:2010}, detector-control attack~\cite{Lars:nature:2010,yuan2011resilience, lydersen2011comment,yuan2011reply, Gerhardt:2010}, detector dead-time attack \cite{weier2011quantum} and others~\cite{jain2011device,li2011attacking}. These attacks suggest that quantum hacking has become a major problem for the real-life security of QKD. Although device-independent QKD~\cite{mayers2004self, acin2007device} offers a nearly perfect solution to the quantum hacking problem, it is not practical because it requires near-unity detection efficiency and even then generates an extremely low key rate~\cite{gisin2010proposal}.

In contrast, Lo, Curty and Qi have recently proposed the novel idea of measurement-device-independent QKD (MDI-QKD)~\cite{Lo:MDIQKD}, which not only removes all quantum attacks in the detection part, the most important security loophole of QKD implementations~\cite{Yi:timeshift:2008, Xu:phaseremapping:2010, Lars:nature:2010, Gerhardt:2010,weier2011quantum, jain2011device}, but also offers excellent performance with current technology. Therefore, MDI-QKD has attracted a lot of scientific attention from the research community on both theoretical~\cite{tamaki2012phase, ma2012statistical, wang2013three, Feihu:practical, marcos:finite:2013} and experimental~\cite{Tittel:2012:MDI:exp, Liu:2012:MDI:exp, zhiyuan:experiment:2013} studies.

\begin{figure}[!t]
\centering
\resizebox{7cm}{!}{\includegraphics{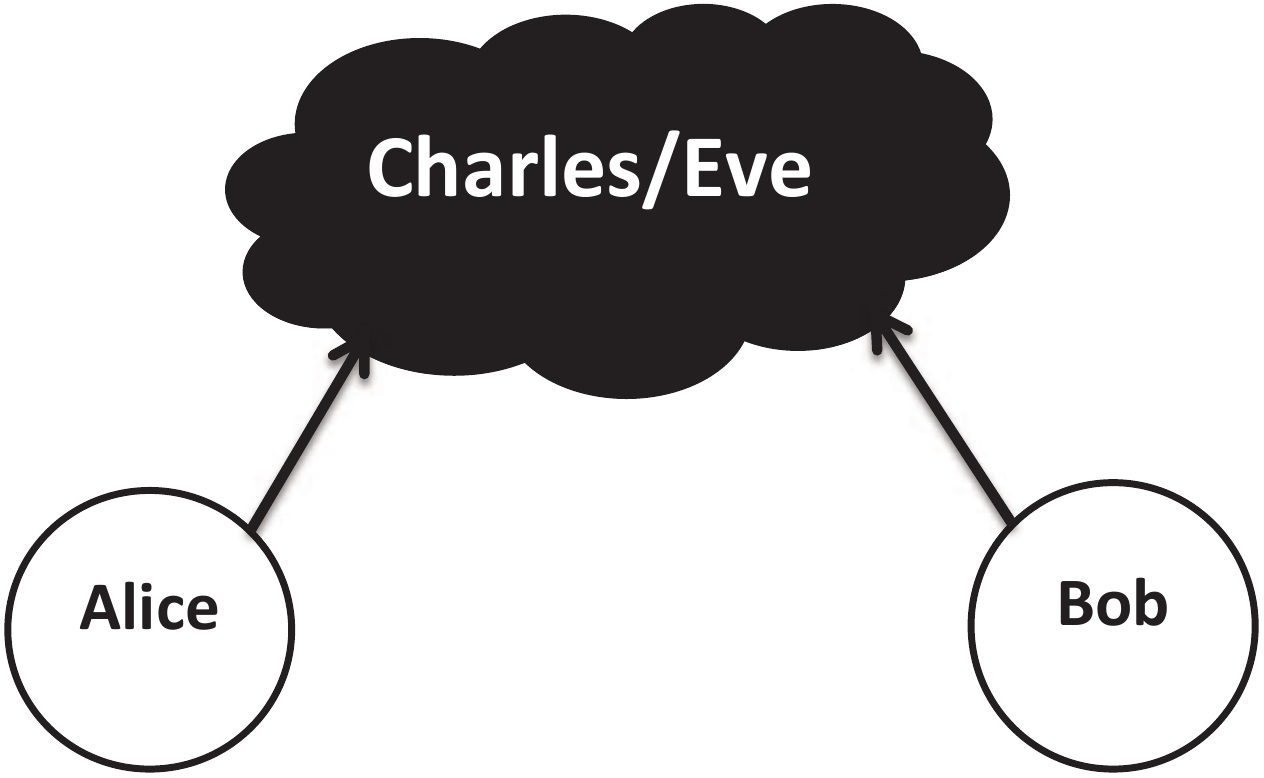}} \caption{\textbf{A general MDI-QKD diagram.} Each of Alice and Bob locally prepares quantum states in the decoy-state BB84 protocol and sends them via a quantum channel to an \emph{untrusted} quantum relay, Charles (or Eve), who performs quantum measurements and broadcasts his/her measurement results. In a general setting, Alice and Bob can use various quantum sources such as a weak-coherent-pulse (WCP) or a parametric-down-conversion (PDC) source, while Charles can perform any general quantum measurement to establish correlations on the information between Alice and Bob. The original MDI-QKD~\cite{Lo:MDIQKD} considers a simple and practically achievable setting where Alice and Bob employ WCP as the source and Charles uses a Bell state measurement (BSM) for quantum measurement. } \label{Fig:General:model}
\end{figure}

In a general MDI-QKD scheme (see Fig.~\ref{Fig:General:model}), each of Alice and Bob locally prepares quantum states, via photons, according to the phase randomized decoy-state BB84 protocol~\cite{Hwang:2003,Lo:2005,Wang:2005} and sends them via a quantum channel to an \emph{untrusted} quantum relay, Charles (or Eve), who is supposed to perform quantum measurements and broadcast his/her measurement results. Since the measurement device in MDI-QKD is essentially used to post-select entanglement between Alice and Bob~\cite{Lo:MDIQKD}, it can be treated as a true black box, which indicates that MDI-QKD is naturally immune to all detection loopholes. This is a major achievement as MDI-QKD allows legitimate users to not only perform secure quantum communications with untrusted relays~\footnote{This also implies the feasibility of ``Pentagon Using China Satellite for U.S.-Africa Command''. See http://www.bloomberg.com/news/2013-04-29/pentagon-using-china-satellite-for-u-s-africa-command.html.} but also out-source the manufacturing of detectors to untrusted manufactures. Notice that detectors are often the most technologically demanding components of a QKD system~\cite{QC:Gisin:2002}, whereas sources can be simple lasers that could be made compact and low-cost~\cite{hughes2013network}.

The original MDI-QKD~\cite{Lo:MDIQKD} considers a simple setting where Alice and Bob employ weak coherent pulses for state preparation and Charles uses a Bell state measurement (BSM) for quantum measurement. It can be easily implemented with standard optical components and thus experimental attempts have been made by several groups~\cite{Tittel:2012:MDI:exp, Liu:2012:MDI:exp, zhiyuan:experiment:2013}. Nevertheless, even in the asymptotic case, the initial protocol can only tolerate a maximal loss of 50 dB, which imposes a limit on the transmission distance within 238km for standard telecom fiber.

In this paper, we propose an important extension of MDI-QKD, called MDI-QKD \emph{with} entangled photon sources in the middle. This method is a detector-loophole-free and ultra-long-distance QKD protocol, which is highly compatible to the future applications of network-centric quantum communications~\cite{hughes2013network}, as the entangled photon sources together with the BSM devices (in the middle) in this method can be totally untrusted. In particular, we analyze one specific case in which the quantum relay is composed of one entangled photon source and two BSMs. This scheme is simple, experimentally feasible, and most importantly, it enables us to implement QKD over significantly longer distance than what is possible with any existing proposal in the literature.

\begin{figure}[!t]
\centering
\resizebox{7cm}{!}{\includegraphics{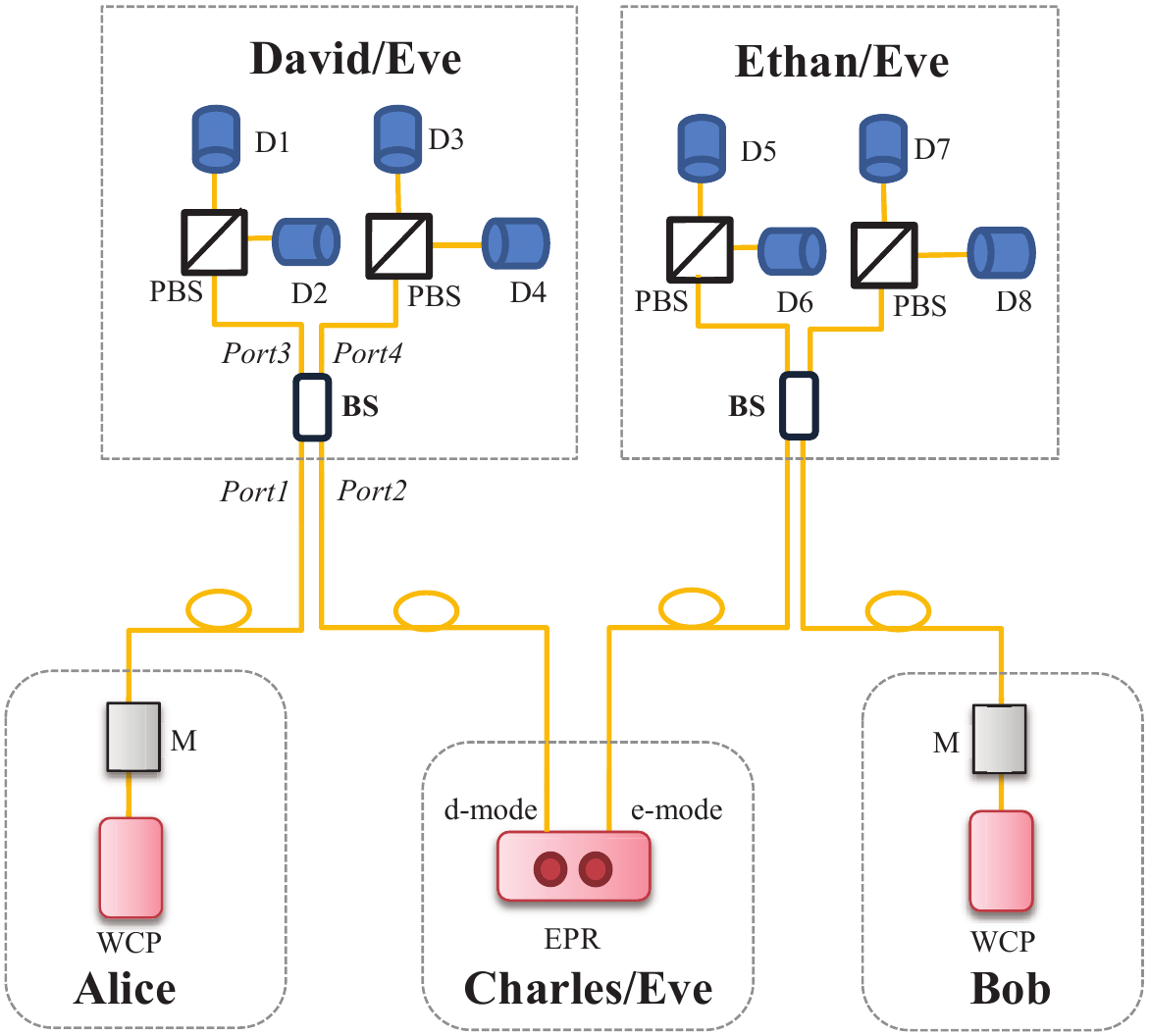}} \caption{\textbf{MDI-QKD with one entangled photon source in the middle.} WCP: weak coherent pulse; EPR: Type II parametric-down-conversion source; M: polarization and intensity modulators; BS: beam splitter; D: single photon detector (SPD); PBS: polarization beam splitter (in the ideal case without any polarization misalignment, the $Z$ basis, used for key generation in Eq.~(\ref{Eqn:Key:formula}), refers to the basis of PBS).} \label{Fig:model}
\end{figure}

Fig.~\ref{Fig:model} illustrates the diagram of MDI-QKD \emph{with} one entangled photon source in the middle. This method is similar to the original MDI-QKD in that David, Ethan and Charles together can be treated as an untrusted relay. Hence, the model and security analysis are nearly equivalent. Note that this analysis is also applicable to the case of multiple entangled photon sources combined with multiple BSMs in the middle.

The protocol is as follows. Each of Alice and Bob prepares phase-randomized weak coherent pulses (WCP) in one of the four BB84 polarization states~\cite{bennett1984quantum} randomly and independently. They also randomly modulate the average photon number in each pulse to implement the decoy-state method~\cite{Hwang:2003,Lo:2005,Wang:2005}. Meanwhile, an \emph{untrusted} source, Charles, prepares polarization entangled photon pairs using a Type II parametric-down-conversion (PDC) source (ideally, producing Singlet $|\psi^{-}\rangle$=$\frac{1}{\sqrt{2}}(|H,V\rangle-|V,H\rangle$). All three parties send quantum signals to two untrusted relays, David and Ethan, each of whom is supposed to perform a BSM that projects the incoming signals into a Bell state (either Singlet $|\psi^{-}\rangle$ or Triplet $|\psi^{+}\rangle$=$\frac{1}{\sqrt{2}}(|H,V\rangle+|V,H\rangle$). Here, Alice and Bob can use the decoy-state method to estimate the single-photon contributions~\cite{ma2012statistical, wang2013three, Feihu:practical, marcos:finite:2013}, \ie, estimate the counts and the error rates when both Alice and Bob send out single-photon pulses and both David and Ethan report successful events.

In the classical communication phase, each of David and Ethan uses a classical channel to broadcast their measurement results. Alice and Bob keep the successful events (\ie, the event when both David and Ethan achieve successful BSMs), discard the rest and post-select the events where they use the same basis. Finally, as shown in Table~\ref{tab:BSM}, either Alice or Bob applies a bit flip to her or his data according to their basis and the BSM results.

For post-processing, Alice and Bob evaluate the data sent in two bases separately~\cite{lo2005efficient}. The \emph{Z}-basis (rectilinear) is used for key generation, while the \emph{X}-basis (diagonal) is used for testing against tampering and the purpose of quantifying the amount of privacy amplification needed. In the \emph{Z}-basis, an error corresponds to a successful event when Alice and Bob prepare the same quantum states; in the \emph{X}-basis, an error corresponds to a projection into $|\psi^{-}\psi^{-}\rangle$ or $|\psi^{+}\psi^{+}\rangle$ when they prepare the same states, or, into $|\psi^{+}\psi^{-}\rangle$ or $|\psi^{-}\psi^{+}\rangle$ when they prepare orthogonal states (see Table~\ref{tab:BSM}). The secure key rate in the asymptotic case (\ie, with an infinite number of transmission signals and decoy states) is given by~\cite{Lo:MDIQKD}
\begin{equation} \label{Eqn:Key:formula}
    R=Q_{Z}^{1,1}[1-H_{2}(e_{X}^{1,1})]-Q_{Z}f_{e}(E_{Z})H_{2}(E_{Z})
\end{equation}
where $Q_{Z}$ and $E_{Z}$ denote, respectively, the gain and quantum bit error rate (QBER) in the $Z$ basis; $f_{e}\geq 1$ is the error correction inefficiency function and in this paper, we assume $f_{e}=1.16$; $Q_{Z}^{1,1}$ and $e_{X}^{1,1}$ are the gain and error rate when both Alice and Bob send single-photon states. In practice, $Q_{Z}$ and $E_{Z}$ are directly measured from experiments, while $Q_{Z}^{1,1}$ and $e_{X}^{1,1}$ can be estimated from the finite decoy-state method~\cite{Feihu:practical, marcos:finite:2013}.

To evaluate the performance of our protocol, we provide a general approach to model the system. Although the model is proposed to study MDI-QKD, it is also useful for other non-QKD experiments involving entanglement and BSMs~\cite{kok2007linear,sangouard2011quantum}. In this model, the source is a composite of two weak coherent states prepared by Alice and Bob and one EPR state (Singlet) prepared by Charles; The polarization rotations (\ie, polarization misalignments) and losses of the transmissions of the four quantum channels (\ie, Alice to David, Bob to Ethan, Charles to David and Ethan) are respectively modelled by four unitary matrices~\cite{kok2007linear} and four beam-splitters; The measurement is realized by two BSMs, each of which contains a typical Hong-Ou-Mandel interference~\cite{HOM:1987}, followed by threshold detections~\footnote{A threshold detector can only tell whether the input signal is vacuum or non-vacuum.}. Finally, we can derive all the terms in Eq.~(\ref{Eqn:Key:formula}). The details of our model are shown in the supplementary material~\footnote{See supplementary material at [URL will be inserted by AIP] for the model of the system.}.

\begin{figure}[!t]
\centering
\resizebox{7cm}{!}{\includegraphics{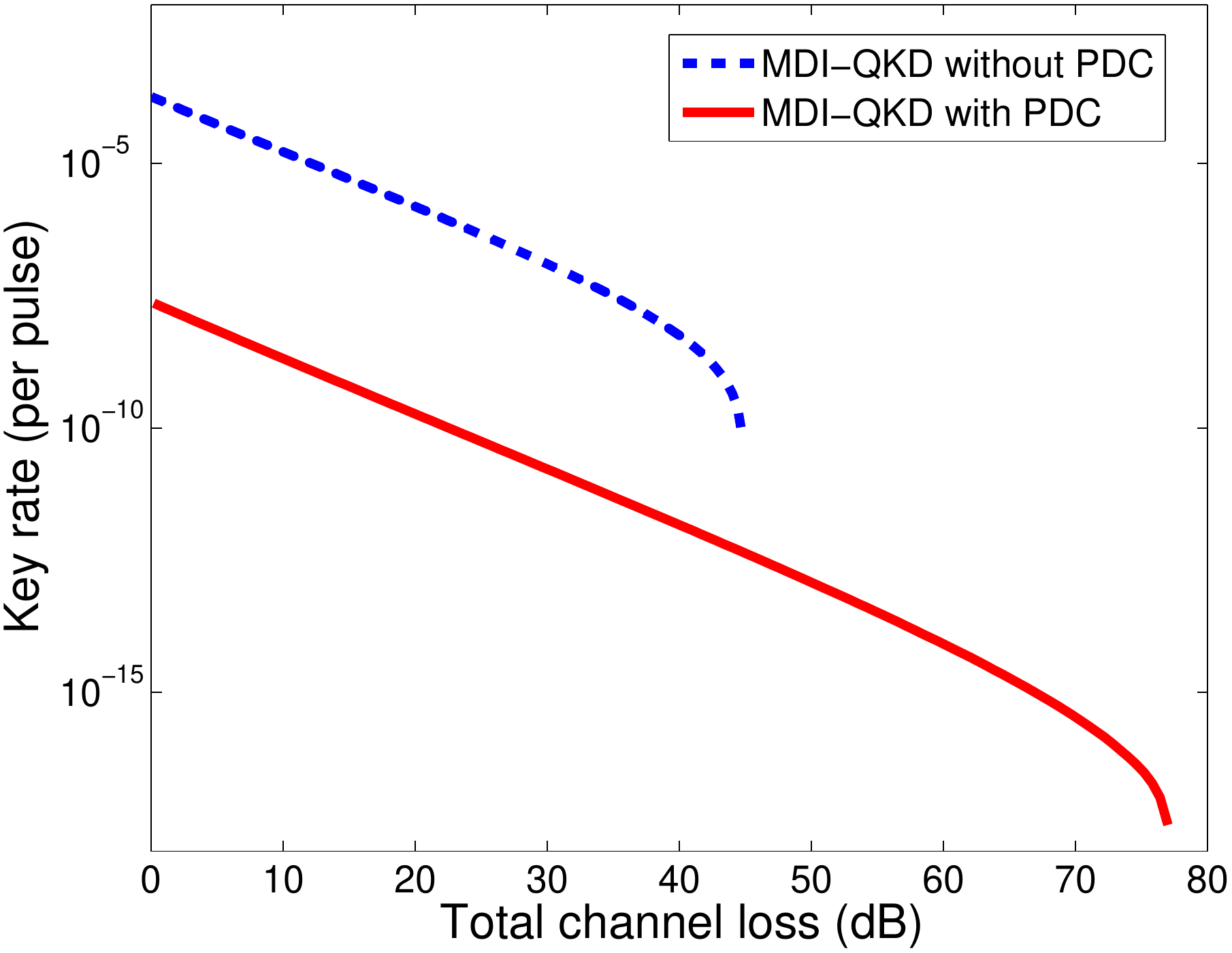}} \caption{\textbf{Asymptotic key rate.} Asymptotic limit means that Alice and Bob have an infinite number of signals and decoy states. The practical parameters are from an entanglement based QKD experiment~\cite{Ursin:144QKD}: the detection efficiency is 14.5\%, the background count rate is $6.02\times 10^{-6}$, and the intrinsic error rate due to misalignment and instability of the system is 3\% (owing to 4-channel links in Fig.~\ref{Fig:model} instead of 2 links in Ref.~\cite{Ursin:144QKD}, the total misalignment error is assumed to be roughly twice as that in Ref.~\cite{Ursin:144QKD}.). In the low and medium channel loss regions, since MDI-QKD \emph{with} one PDC source requires 4-fold coincident detections instead of 2-fold coincident detections required by the original MDI-QKD, its key rate is lower than that of the original MDI-QKD. However, MDI-QKD with one PDC source can tolerate significantly higher losses up to 77 dB (367km standard telecom fiber).} \label{Fig:Key:Aymptotic}
\end{figure}

In simulation, the polarization misalignments of the four quantum channels are assumed to be identical, and the four channel transmittances are optimized by maximizing the key rates. Firstly, we simulate the key rate in the asymptotic case using the practical parameters from the entanglement based QKD experiment reported in Ref.~\cite{Ursin:144QKD}. This result is shown by the red solid curve in Fig.~\ref{Fig:Key:Aymptotic}. As a comparison, we also present the simulation result of the original MDI-QKD~\cite{Lo:MDIQKD} in this figure (see the blue dashed curve). It is interesting that MDI-QKD \emph{with} one PDC source in the middle can tolerate significantly higher loss, up to 77 dB. Notice that with the same practical parameters, the decoy-state BB84 protocol, however, can only tolerate around 30 dB~\cite{ma2007quantum}. Without other losses, a 77 dB loss corresponds to a channel transmission of 367km standard telecom fiber (0.21 dB/km) or 481km ultra-low loss telecom fiber (0.16 dB/km~\cite{ohashi1992optical}).

It is worth noting that in Fig.~\ref{Fig:Key:Aymptotic}, the optimal key rate of MDI-QKD \emph{with} one PDC source at 0km is about $2.39\times10^{-8}$ bits per pulse. Why is this key rate lower than the original MDI-QKD? It is due to two factors: 1) MDI-QKD with one PDC source requires 4-fold coincidence, whereas the original MDI-QKD requires only 2-fold coincidence; Hence, the low detector efficiency here (14.5\%) inherently decreases the key rate by around two orders of magnitude. 2) If the PDC source in the middle presents a large brightness, its multi-photon pairs contribute significantly to the QBER. Consequently, the optimal brightness of this PDC source is on the order of $10^{-3}$.

\begin{figure}[!t]
\centering
\resizebox{7cm}{!}{\includegraphics{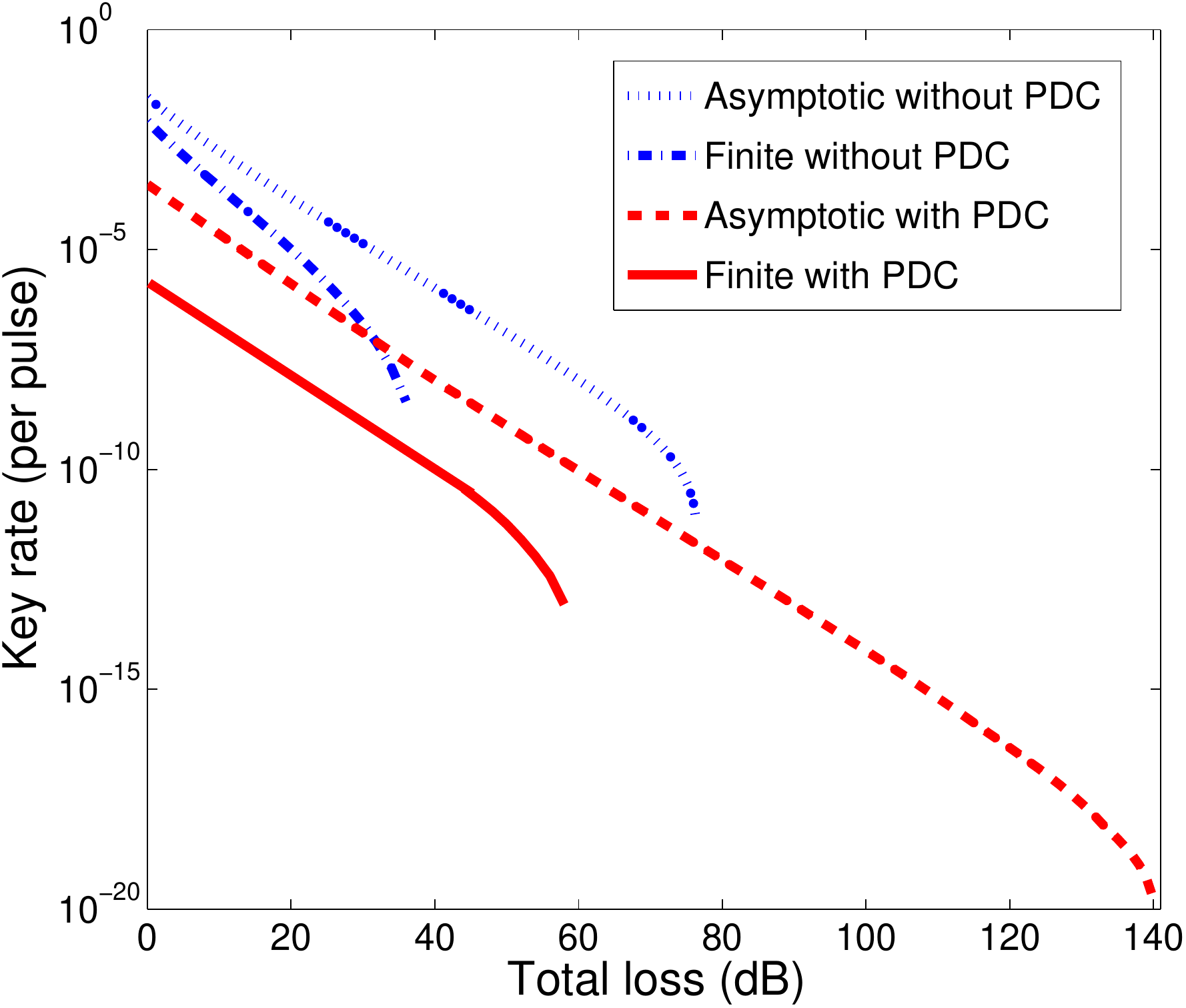}} \caption{\textbf{Key rate with state-of-the-art SPDs.} We consider better SPDs with detecter effiency of 93\% and dark count rate of $1\times 10^{-6}$. The other experimental parameters are the same as those used in Fig.~\ref{Fig:Key:Aymptotic}. In the asymptotic limit, MDI-QKD with one PDC source can tolerate significantly higher losses, up to 140 dB (667km standard fiber). With the method of Ref.~\cite{Feihu:practical,marcos:finite:2013}, the finite-key analysis is conducted on a data-size of $N$=$10^{15}$. MDI-QKD with one PDC source can tolerate 60 dB loss (286km standard fiber) in the finite-key case.} \label{Fig:Key:Aymptotic:betterDet}
\end{figure}

The result in Fig.~\ref{Fig:Key:Aymptotic} can be significantly improved if we consider state-of-the-art single-photon detectors (SPD). For instance, using the practical parameters of Ref.~\cite{marsili2013detecting} with detector efficiency of 93\% and dark-count rate (per gating window) of $1\times 10^{-6}$, the simulation result is shown in Fig.~\ref{Fig:Key:Aymptotic:betterDet}. Remarkably, it shows that our scheme can tolerate 140 dB loss (667km standard fiber) in the asymptotic limit. The optimal brightness of PDC is still on the order of $10^{-3}$. To address the finite-key effect, we simulate the finite-key rate by blue dash-dotted and red solid curves in this figure. Here we use the method reported in Ref.~\cite{marcos:finite:2013} for a rigorous finite-key analysis including an analytical approach~\cite{Feihu:practical} with two decoy states for the finite decoy-state protocol, a data-size of $N$=$10^{15}$ and a security bound of $\epsilon$=$10^{-10}$ for the finite-key analysis. In this case, our scheme can tolerate up to 60 dB channel loss (286km standard fiber or 375km ultra-low loss fiber).

The ultimate transmission distance is limited by the low system operation rate, \ie, the speed of experimental devices. For instance, at 100 dB, even in the asymptotic case, the optimal key rate is only $3\times10^{-15}$. Thus, to get one secure bit requires 30 hours of continuous experiment with a high-speed QKD system working at 10 GHz. Note that achieving such high-speed system is currently one of the primary goals of experimental quantum communication community. On the other hand, similar to classical optical communication, a quantum repeater~\cite{sangouard2011quantum} can be helpful for an ultra-long distance quantum communication.

In summary, we have presented a feasible method of MDI-QKD with entangled photon sources in the middle. This method is simple, experimentally feasible, and most importantly, it enables us to implement detection-loophole-free QKD over ultra-long distances. Our work is relevant to not only QKD but also general experiments involving entangled photon sources and BSMs. \\

We thank enlightening discussions with M.~Curty, W.~Cui, S. Gao, D.~Kang, L.~Qian, C.~Weedbrook and F.~Ye. Support from funding agencies NSERC and the CRC program is gratefully acknowledged. F. Xu would like to thank the Paul Biringer Graduate Scholarship for the financial support.

\begin{table}[!h]\center
\caption{Successful Bell state measurements.}\label{tab:BSM}
\begin{tabular}{c @{\hspace{0.1cm}} c @{\hspace{0.15cm}} c @{\hspace{0.15cm}} c @{\hspace{0.15cm}} c @{\hspace{0.15cm}} c}
\hline
David\&Ethan &  & $|\psi^{+}\psi^{+}\rangle$ & $|\psi^{-}\psi^{-}\rangle$ & $|\psi^{+}\psi^{-}\rangle$ & $|\psi^{-}\psi^{+}\rangle$\\
\hline
Alice/Bob  & \emph{Z}-basis & flip & flip & flip & flip\\
Alice/Bob  & \emph{X}-basis & flip & flip & non-flip & non-flip\\
\hline
\end{tabular}
\end{table}

\appendix~\label{App:model}

\section{Model}
We discuss our general approach to model the system of MDI-QKD \emph{with} one PDC source in the middle. The asymptotic key rate is given by
\begin{equation} \label{Eqn:Key:formula}
    R=Q_{Z}^{1,1}[1-H_{2}(e_{X}^{1,1})]-Q_{Z}f_{e}(E_{Z})H_{2}(E_{Z})
\end{equation}
In what follows, we discuss how one can derive each quantity in this key rate formula, \ie, $Q_{Z}^{1,1}$, $e_{X}^{1,1}$, $Q_{Z}$ and $E_{Z}$.

\subsection{Preliminary}
\begin{description}
  \item[Notations:] $\mu_{a}$ ($\mu_{b}$) is the mean photon number of Alice's (Bob's) coherent state; $\mu_{c}$ is the expected photon pairs of Charlie's PDC source; $\mu_{a}^{opt}$ ($\mu_{b}^{opt}$, $\mu_{c}^{opt}$) is the optimized $\mu$ that maximizes the key rate; $L_{ad}$ and $t_{a}$ ($L_{be}$ and $t_{b}$) denote, respectively, the channel distance and transmittance from Alice (Bob) to David (Ethan). For a fiber-based system, $t_{a}$=$10^{-\alpha L_{ad}/10}$ with $\alpha$ denoting the channel loss coefficient; Similarly, $L_{cd}$ and $t_{d}$ ($L_{ce}$ and $t_{e}$) denote the ones from Charles to David (Ethan); $\eta_{d}$ is the detector efficiency and $Y_{0}$ is the dark-count rate; $e_{d}$ denotes the \emph{total} polarization misalignment error.

  \item[Misalignment:] In our simulation, for simplicity, we consider a 2-dimensional unitary matrix to represent the polarization rotation (or misalignment) of each channel transmission. Notice that this unitary matrix is a simple form rather than the general one~\cite{kok2007linear}. Nonetheless, we believe that the result for a more general unitary transformation will be similar to our simulation results. This unitary matrix is given by $U_{k}$=$\left(
  \begin{array}{cc}
    \cos\theta_{k} & -\sin\theta_{k} \\
    \sin\theta_{k} & \cos\theta_{k} \\
  \end{array}
\right)$ with $k\in\{1,2,3,4\}$. $\{U_{1},U_{2},U_{3},U_{4}\}$ denotes the misalignments for the channels of \{$L_{ad}$, $L_{cd}$, $L_{ce}$, $L_{be}$\} respectively. The misalignment error, $e_{dk}$, is defined as $e_{dk}=\sin^{2}\theta_{k}$. Since the misalignment error is relatively independent of the channel distance, we assume that $e_{dk}$ is \emph{equal} with each other, thus $e_{dk}$=$\frac{e_{d}}{4}$. We also assume that $\theta_{k}$ is randomly distributed in $\{-\theta_{k}^{max},+\theta_{k}^{max}\}$ with $\theta_{k}^{max}=\arcsin(\sqrt{e_{dk}})$. We remark that our model is a simple extension of our previous model~\cite{Feihu:practical} used in MDI-QKD without entangled photon sources.

\end{description}

\subsection{Source}
\emph{WCP:} The output from an attenuated laser is a weak-coherent state $\ket{\alpha}$ that is a superposition of number states (Fock states). Assuming that the phase of the laser is totally randomized for each pulse, the photon number of each pulse follows a Poison distribution with a parameter $\mu$=$|\alpha|^2$ as its mean photon number. Hence, the density matrix of the weak-coherent state is given by
\begin{equation}\label{Model:CoherentState}
\begin{aligned}
\rho = \frac{1}{2\pi} \int\limits_{0}^{2\pi} d \theta \ket{ |\alpha| e^{i\theta} } \bra{ |\alpha| e^{i\theta} }\\
= \sum^{\infty}_{n=0}\frac{\mu^n}{n!}e^{-\mu} \ket{n}\bra{n}
\end{aligned}
\end{equation}
where $\theta$ is the phase of the state and $\ket{n}\bra{n}$ is the density matrix of the $n$-photon state.

\emph{EPR:} The state emitted from a type-II PDC source can be written as~\cite{kok2001detection}
\begin{equation}\label{Eqn:Model:PDCstate}
\begin{aligned}
|\Psi\rangle=(\cosh\chi)^{-2}\sum_{n=0}^{\infty}\sqrt{n+1}\tanh^n\chi|\Phi_n\rangle,
\end{aligned}
\end{equation}
where $|\Phi_n\rangle$ is the state of an $n$-photon pair, given by
\begin{equation}\label{Eqn:Model:PDCn}
\nonumber
\begin{aligned}
|\Phi_n\rangle=\frac{1}{\sqrt{n+1}}\sum_{m=0}^{n}(-1)^m|n-m,m\rangle_d\otimes|m,n-m\rangle_e.
\end{aligned}
\end{equation}
The probability of an $n$-photon pair is $P(n)$=$\frac{(n+1)\lambda^n}{(1+\lambda)^{n+2}}$, where $\lambda$=$\sinh^2\chi$. The expected number (brightness) of photon pairs per pump pulse is $\mu$=$2\lambda$.

Here, we use the \emph{polarization} modes as the qubit basis. Specifically, in \emph{Z}-basis, $|m,n\rangle$ represents $m$ photons in $|H\rangle$ mode (horizontal) and $n$ photons in $|V\rangle$ mode (vertical); while in \emph{X}-basis, $|m,n\rangle$ represents $m$ photons in $|+\rangle$ mode (45) and $n$ photons in $|-\rangle$ mode (135).
\subsection{Transmission}
Suppose Alice, Bob, and Charles send out states $\ket{\alpha_a}$, $\ket{\alpha_b}$ (Eq.~(\ref{Model:CoherentState})) and $\ket{\Psi_c}$ (Eq.~(\ref{Eqn:Model:PDCstate})) respectively. After channel transmission, the source states evolve to
\begin{eqnarray}\label{Eqn:coherent:EPR}
\ket{\alpha_a}\rightarrow\ket{\alpha'_a}=\sum_{x_a=0}^{\infty}\sum_{y_a=0}^{\infty}C_a|x_a, y_a\rangle \nonumber \\
\ket{\alpha_b}\rightarrow\ket{\alpha'_b}=\sum_{x_b=0}^{\infty}\sum_{y_b=0}^{\infty}C_b|x_b, y_b\rangle \nonumber \\
\ket{\Psi_c}\rightarrow\ket{\phi_c}=\sum_{x_d=0}^{\infty}\sum_{y_d=0}^{\infty}\sum_{x_e=0}^{\infty}\sum_{y_e=0}^{\infty}C_c|x_d, y_d, x_e, y_e\rangle
\end{eqnarray}
where $x$ and $y$ denote the number of photons in $|H\rangle$ and $|V\rangle$ mode respectively and $C$ denotes the final coefficient associated with channel loss and misalignment. In the following, we discuss how one can derive $C_a$, $C_b$ and $C_c$.

\emph{WCP:} Suppose Alice and Bob send out coherent states both in $|H\rangle$ mode, after channel transmission with transmittance \{$t_a$, $t_b$\} and misalignment angle \{$\theta_1$, $\theta_4$\}, the resulting density matrices can be written as
\begin{eqnarray} \label{Eqn:coherent:loss}
\ket{\rho'_a}=e^{-{\mu_{a}t_a}}\sum_{n_a=0}^{\infty}\sum_{m_a=0}^{n_a}\frac{(\mu_{a}t_a)^{n_a}}{n_{a}!}{n_a\choose m_a}(\cos^2\theta_1)^{m_a} \times \nonumber \\
(\sin^2\theta_1)^{(n_a-m_a)}\ket{m_a,n_a-m_a}\bra{m_a,n_a-m_a} \nonumber \\
\ket{\rho'_{b}}=e^{-{\mu_{b}t_b}}\sum_{n_b=0}^{\infty}\sum_{m_b=0}^{n_b}\frac{(\mu_{b}t_b)^{n_b}}{n_{b}!}{n_b\choose m_b}(\cos^2\theta_4)^{m_b} \times \nonumber \\
(\sin^2\theta_4)^{n_b-m_b}\ket{m_b, n_b-m_b}\bra{m_b, n_b-m_b} \nonumber
\end{eqnarray}
where $m$ ($n$-$m$) denotes the number of photons in $|H\rangle$ ($|V\rangle$) mode. Hence, we can derive the coefficients $C_a$ and $C_b$ (Eq.~(\ref{Eqn:coherent:EPR})).

\emph{EPR:} Let us start from a general state $|\Psi\rangle$ emitted by Charles. After channel transmission \{$t_d$, $t_e$\}, the resulting state is
\begin{eqnarray} \label{Eqn:EPR:loss}
(\cosh\chi)^{-2}\sum_{n_c=0}^{\infty}\tanh^{n_c}\chi\sum_{m_c=0}^{n_c}(-1)^{m_c}\big\{ [ \nonumber \\
\sum_{k_d=0}^{m_c}\sum_{(l_d-k_d)=0}^{n_c-m_c} \sqrt{\binom {n_c-m_c} {l_d-k_d}{m_c\choose k_d}} \times \nonumber \\
(\sqrt{t_d})^{l_d}(\sqrt{1-t_d})^{n_c-l_d}|l_d-k_d, k_d\rangle_d] \nonumber \\
\otimes[ \sum_{k_e=0}^{m_c}\sum_{(l_e-k_e)=0}^{n_c-m_c}\sqrt{\binom {n_c-m_c}{l_e-k_e}{m_c\choose k_e}} \times \nonumber \\
(\sqrt{t_e})^{l_e}(\sqrt{1-t_e})^{n_c-l_e}|k_e, l_e-k_e\rangle_e] \big\}
\end{eqnarray}
where $l_d$ and $l_e$ denote the number of photons passing the channel and finally arriving at David and Ethan. Afterwards, combined with channel misalignment \{$\theta_2$, $\theta_3$\}, the above joint state, \ie, $|l_d-k_d, k_d\rangle_d\otimes|k_e, l_e-k_e\rangle_e$, is given by
\begin{eqnarray} \label{Eqn:EPR:misalignment}
\sum_{i_d=0}^{k_d}\sum_{j_d=0}^{l_d-k_d}\sqrt{\binom {l_d-k_d} {j_d}{k_d\choose i_d}}(-1)^{l_d-k_d-j_d} \times \nonumber \\
(\cos\theta_2)^{k_d+j_d-i_d}(\sin\theta_2)^{l_d+i_d-k_d-j_d}|i_d+j_d, l_d-i_d-j_d\rangle_d \nonumber \\
\otimes \sum_{i_e=0}^{k_e}\sum_{j_e=0}^{l_e-k_e}\sqrt{\binom {l_e-k_e} {j_e}{k_e\choose i_e}}(-1)^{k_e-i_e} \times \nonumber \\
(\cos\theta_3)^{i_e+l_e-k_e-j_e}(\sin\theta_3)^{k_e+j_e-i_e}|i_e+j_e, l_e-i_e-j_e\rangle_e
\end{eqnarray}

Finally, by combining Eqs.~(\ref{Eqn:EPR:loss}) and (\ref{Eqn:EPR:misalignment}), we can derive the coefficient $C_c$ in Eq.~(\ref{Eqn:coherent:EPR}). In simulation, we assume that the channel transmittances and the polarization misalignments of the 4 channels are identical~\footnote{One might ask `whether it is possible to increase the key rate by considering \emph{unequal} transmissions, \ie, $t_a>t_d$ and $t_b>t_e$?' Because, in such case, one can enhance the brightness of PDC regardless of multi-photon pairs (increasing QBER) and thus improve the key rate. However, we found that unequal transmissions could not increase the key rate too much. The key reason is that when $t_a>t_d$, the multi-photon pulse of WCP combined with the channel misalignment of $L_{ad}$ will take turns to contribute significantly to the QBER. In our simulation, we simultaneously optimize the channel lengths (\{$L_{ad}$, $L_{cd}$\} and \{$L_{be}$, $L_{ce}$\}) and the brightness of WCP and PDC, and finally simulate the optimal key rates shown by the curves in the main text.}
\subsection{Detection}
$\ket{\alpha'_a}$, $\ket{\alpha'_b}$, and $\ket{\phi_c}$ will finally interfere (Hong-Ou-Mandel interference~\cite{HOM:1987}) at David and Ethan. In Eq.~(\ref{Eqn:coherent:EPR}), each one is a superposition of number states. Hence, the overall HOM interference is the \emph{superposition} of the interference between all number states.

Let us focus on one specific input number state (of BS), interfering at David and Ethan:
\begin{equation}\label{Eqn:HOM:General}
\begin{aligned}
\big\{|x_a, y_a\rangle\otimes|x_d, y_d\rangle\big\}_{David}^{in} \otimes \big\{|x_e, y_e\rangle\otimes|x_b, y_b\rangle\big\}_{Ethan}^{in}
\end{aligned}
\end{equation}

In $Z$ basis, considering $|H\rangle$ and $|V\rangle$ mode separately, we can derive the interference results (output of the two beam splitters) using the method of Ref.~\cite{Rarity:2005:HOM}. For instance, on David's side, the interference for $|H\rangle$ mode is between $|x_a\rangle_{port1-h}$ and $|x_d\rangle_{port2-h}$, where the interference result is given by a binomial distribution
\begin{equation}\label{Eqn:HOM:HH}
\nonumber
\begin{aligned}
\sum_{q_a=0}^{x_a}\sum_{q_d=0}^{x_d}{x_a\choose q_a}\binom {x_d}{q_d} \sqrt{\frac{(q_a+q_d)!}{x_a!x_d!}} \times \nonumber \\
\sqrt{(x_a+x_d-q_a-q_d)!}t^{(q_a+x_d-q_d)}r^{(x_a-q_a+q_d)} \nonumber \\
|q_a+q_d\rangle_{port3-h}\otimes|x_a+x_d-q_a-q_d\rangle_{port4-h}
\end{aligned}
\end{equation}
where $t$ ($r$) is the transmission (reflection) coefficient of BS satisfying $r^{*}t+rt^{*}$=0 and $|r|^2+|t|^2$=1. Hence, the coefficient of $z$ photons in $|H\rangle$ mode populating out from port3 of David's BS, $C_{port3-h}(z)$, is given by
\begin{equation}\label{Eqn:HOM:HH}
\nonumber
\begin{aligned}
\sum_{k=0}^{x_a}{x_a\choose k}\binom {x_d}{z-k}\sqrt{\frac{z!(x_a+x_d-z)!}{x_a!x_d!}} \nonumber \\
t^{(2k+x_d-z)}r^{(x_a+z-2k)}
\end{aligned}
\end{equation}
where $k\leq z$, $(z-k)\leq (x_d)$, and $z\in\{0,1,...,(x_a+x_d)\}$. Note that $C_{port3-h}(z)$ is also the coefficient of $x_a+x_d-z$ photons populating out from port4 of David's BS. Similarly, we can get the interference result for $|V\rangle$ mode, \ie, between $|y_a\rangle_{port1-v}$ and $|y_d\rangle_{port2-v}$.

At the same time, we can also derive the interference result between $|x_e, y_e\rangle_{c}$ and $|x_b, y_b\rangle_{b}$ on Ethan's side and thus the joint interference result (4-fold coincidence) for a given number state given by Eq.~(\ref{Eqn:HOM:General}).

By summing over all the number states given by Eq.~(\ref{Eqn:coherent:EPR}), we can calculate the overall inference results. In our simulation, for each number state given by Eq.~(\ref{Eqn:HOM:General}), we create a table to store the coefficients of different interference outputs. By adding the tables for all number states (Eq.~(\ref{Eqn:coherent:EPR})), we can have the summation table containing the final coefficients of all interference outputs. In the end, we can have the coincident detection probabilities by considering the detection efficiency $\eta_d$ of a threshold SPD~\footnote{Notice that for free-space transmission in visible wavelength, silicon single-photon detector can be used, which have higher efficiency, typically over 50\%.}.


\subsection{Key rate}
Based on the detection probabilities, we can derive the gains \ie, $Q^{HH}_{Z}$ and $Q^{HV}_{Z}$, for different encodings by Alice/Bob. Therefore, the overall gain and QBER in $Z$ basis are given by
\begin{equation} \label{Qrect:Erect}
\nonumber
\begin{aligned}
    Q_{Z}&=\frac{Q_{Z}^{HH}+Q_{Z}^{HV}}{2}\\
    E_{Z}&=\frac{Q_{Z}^{HH}}{Q_{Z}^{HH}+Q_{Z}^{HV}}
\end{aligned}
\end{equation}

In the asymptotic case, $Q_Z^{1,1}$ is given by $P^{1,1}Y^{1,1}_{Z}$, where $P^{1,1}$=$\mu_a\mu_b e^{-(\mu_a+\mu_b)}$ denotes the single-single photon probability and $Y_{Z}^{1,1}$ denotes the yield, \ie, the conditional probability that a successful event happens in $Z$ basis when Alice/Bob send single-photon state, which can be derived by substituting Eq.~(\ref{Model:CoherentState}) by a perfect single-photon source. In the finite decoy-state case~\cite{ma2005practical, wang2005decoy}, $Q_Z^{1,1}$ is estimated from the quantifies of $Q_{Z}$ for different intensities. A similar discussion holds when Alice and Bob use X basis for encoding, \ie, $|+\rangle$$|+\rangle$ and $|+\rangle$$|-\rangle$. Thus, we can have the overall gain and QBER for $X$ basis. In the asymptotic case, $e_X^{1,1}$ can be calculated from $Y_{X}^{1,1}$ (the yield in $X$ basis), while in the finite decoy-state case, it is estimated by $Q_{X}$ and $E_{X}$. Finally, we can derive the key rate $R$ given by Eq.~(\ref{Eqn:Key:formula}).

Here in the simulation of the key rate with the finite decoy states (see Fig.4 in the main text), we consider an analytical approach with one signal state and two decoy states using the method reported in Refs.~\cite{Feihu:practical, marcos:finite:2013}. The intensities of the signal and decoy states are optimized by maximizing the key rates.

\bibliographystyle{apsrev4-1}
\bibliography{BibentangledMDI1}

\begin{thebibliography}{59}%
\makeatletter
\providecommand \@ifxundefined [1]{%
 \@ifx{#1\undefined}
}%
\providecommand \@ifnum [1]{%
 \ifnum #1\expandafter \@firstoftwo
 \else \expandafter \@secondoftwo
 \fi
}%
\providecommand \@ifx [1]{%
 \ifx #1\expandafter \@firstoftwo
 \else \expandafter \@secondoftwo
 \fi
}%
\providecommand \natexlab [1]{#1}%
\providecommand \enquote  [1]{``#1''}%
\providecommand \bibnamefont  [1]{#1}%
\providecommand \bibfnamefont [1]{#1}%
\providecommand \citenamefont [1]{#1}%
\providecommand \href@noop [0]{\@secondoftwo}%
\providecommand \href [0]{\begingroup \@sanitize@url \@href}%
\providecommand \@href[1]{\@@startlink{#1}\@@href}%
\providecommand \@@href[1]{\endgroup#1\@@endlink}%
\providecommand \@sanitize@url [0]{\catcode `\\12\catcode `\$12\catcode
  `\&12\catcode `\#12\catcode `\^12\catcode `\_12\catcode `\%12\relax}%
\providecommand \@@startlink[1]{}%
\providecommand \@@endlink[0]{}%
\providecommand \url  [0]{\begingroup\@sanitize@url \@url }%
\providecommand \@url [1]{\endgroup\@href {#1}{\urlprefix }}%
\providecommand \urlprefix  [0]{URL }%
\providecommand \Eprint [0]{\href }%
\providecommand \doibase [0]{http://dx.doi.org/}%
\providecommand \selectlanguage [0]{\@gobble}%
\providecommand \bibinfo  [0]{\@secondoftwo}%
\providecommand \bibfield  [0]{\@secondoftwo}%
\providecommand \translation [1]{[#1]}%
\providecommand \BibitemOpen [0]{}%
\providecommand \bibitemStop [0]{}%
\providecommand \bibitemNoStop [0]{.\EOS\space}%
\providecommand \EOS [0]{\spacefactor3000\relax}%
\providecommand \BibitemShut  [1]{\csname bibitem#1\endcsname}%
\let\auto@bib@innerbib\@empty
\bibitem [{\citenamefont {Kimble}(2008)}]{kimble2008quantum}%
  \BibitemOpen
  \bibfield  {author} {\bibinfo {author} {\bibfnamefont {H.}~\bibnamefont
  {Kimble}},\ }\href@noop {} {\bibfield  {journal} {\bibinfo  {journal}
  {Nature}\ }\textbf {\bibinfo {volume} {453}},\ \bibinfo {pages} {1023}
  (\bibinfo {year} {2008})}\BibitemShut {NoStop}%
\bibitem [{\citenamefont {Kok}\ \emph {et~al.}(2007)\citenamefont {Kok},
  \citenamefont {Munro}, \citenamefont {Nemoto}, \citenamefont {Ralph},
  \citenamefont {Dowling},\ and\ \citenamefont {Milburn}}]{kok2007linear}%
  \BibitemOpen
  \bibfield  {author} {\bibinfo {author} {\bibfnamefont {P.}~\bibnamefont
  {Kok}}, \bibinfo {author} {\bibfnamefont {W.~J.}\ \bibnamefont {Munro}},
  \bibinfo {author} {\bibfnamefont {K.}~\bibnamefont {Nemoto}}, \bibinfo
  {author} {\bibfnamefont {T.~C.}\ \bibnamefont {Ralph}}, \bibinfo {author}
  {\bibfnamefont {J.~P.}\ \bibnamefont {Dowling}}, \ and\ \bibinfo {author}
  {\bibfnamefont {G.}~\bibnamefont {Milburn}},\ }\href@noop {} {\bibfield
  {journal} {\bibinfo  {journal} {Reviews of Modern Physics}\ }\textbf
  {\bibinfo {volume} {79}},\ \bibinfo {pages} {135} (\bibinfo {year}
  {2007})}\BibitemShut {NoStop}%
\bibitem [{\citenamefont {Yin}\ \emph {et~al.}(2012)\citenamefont {Yin},
  \citenamefont {Ren}, \citenamefont {Lu}, \citenamefont {Cao}, \citenamefont
  {Yong}, \citenamefont {Wu}, \citenamefont {Liu}, \citenamefont {Liao},
  \citenamefont {Zhou}, \citenamefont {Jiang} \emph {et~al.}}]{yin2012quantum}%
  \BibitemOpen
  \bibfield  {author} {\bibinfo {author} {\bibfnamefont {J.}~\bibnamefont
  {Yin}}, \bibinfo {author} {\bibfnamefont {J.-G.}\ \bibnamefont {Ren}},
  \bibinfo {author} {\bibfnamefont {H.}~\bibnamefont {Lu}}, \bibinfo {author}
  {\bibfnamefont {Y.}~\bibnamefont {Cao}}, \bibinfo {author} {\bibfnamefont
  {H.-L.}\ \bibnamefont {Yong}}, \bibinfo {author} {\bibfnamefont {Y.-P.}\
  \bibnamefont {Wu}}, \bibinfo {author} {\bibfnamefont {C.}~\bibnamefont
  {Liu}}, \bibinfo {author} {\bibfnamefont {S.-K.}\ \bibnamefont {Liao}},
  \bibinfo {author} {\bibfnamefont {F.}~\bibnamefont {Zhou}}, \bibinfo {author}
  {\bibfnamefont {Y.}~\bibnamefont {Jiang}},  \emph {et~al.},\ }\href@noop {}
  {\bibfield  {journal} {\bibinfo  {journal} {Nature}\ }\textbf {\bibinfo
  {volume} {488}},\ \bibinfo {pages} {185} (\bibinfo {year}
  {2012})}\BibitemShut {NoStop}%
\bibitem [{\citenamefont {Ma}\ \emph {et~al.}(2012{\natexlab{a}})\citenamefont
  {Ma}, \citenamefont {Herbst}, \citenamefont {Scheidl}, \citenamefont {Wang},
  \citenamefont {Kropatschek}, \citenamefont {Naylor}, \citenamefont
  {Wittmann}, \citenamefont {Mech}, \citenamefont {Kofler}, \citenamefont
  {Anisimova} \emph {et~al.}}]{ma2012quantum}%
  \BibitemOpen
  \bibfield  {author} {\bibinfo {author} {\bibfnamefont {X.-S.}\ \bibnamefont
  {Ma}}, \bibinfo {author} {\bibfnamefont {T.}~\bibnamefont {Herbst}}, \bibinfo
  {author} {\bibfnamefont {T.}~\bibnamefont {Scheidl}}, \bibinfo {author}
  {\bibfnamefont {D.}~\bibnamefont {Wang}}, \bibinfo {author} {\bibfnamefont
  {S.}~\bibnamefont {Kropatschek}}, \bibinfo {author} {\bibfnamefont
  {W.}~\bibnamefont {Naylor}}, \bibinfo {author} {\bibfnamefont
  {B.}~\bibnamefont {Wittmann}}, \bibinfo {author} {\bibfnamefont
  {A.}~\bibnamefont {Mech}}, \bibinfo {author} {\bibfnamefont {J.}~\bibnamefont
  {Kofler}}, \bibinfo {author} {\bibfnamefont {E.}~\bibnamefont {Anisimova}},
  \emph {et~al.},\ }\href@noop {} {\bibfield  {journal} {\bibinfo  {journal}
  {Nature}\ }\textbf {\bibinfo {volume} {489}},\ \bibinfo {pages} {269}
  (\bibinfo {year} {2012}{\natexlab{a}})}\BibitemShut {NoStop}%
\bibitem [{\citenamefont {Bennett}\ and\ \citenamefont
  {Brassard}(1984)}]{bennett1984quantum}%
  \BibitemOpen
  \bibfield  {author} {\bibinfo {author} {\bibfnamefont {C.~H.}\ \bibnamefont
  {Bennett}}\ and\ \bibinfo {author} {\bibfnamefont {G.}~\bibnamefont
  {Brassard}},\ }in\ \href@noop {} {\emph {\bibinfo {booktitle} {Proceedings of
  IEEE International Conference on Computers, Systems and Signal
  Processing}}},\ Vol.\ \bibinfo {volume} {175}\ (\bibinfo {organization}
  {Bangalore, India},\ \bibinfo {year} {1984})\BibitemShut {NoStop}%
\bibitem [{\citenamefont {Ekert}(1991)}]{ekert1991quantum}%
  \BibitemOpen
  \bibfield  {author} {\bibinfo {author} {\bibfnamefont {A.~K.}\ \bibnamefont
  {Ekert}},\ }\href@noop {} {\bibfield  {journal} {\bibinfo  {journal}
  {Physical Review Letters}\ }\textbf {\bibinfo {volume} {67}},\ \bibinfo
  {pages} {661} (\bibinfo {year} {1991})}\BibitemShut {NoStop}%
\bibitem [{\citenamefont {Gisin}\ \emph {et~al.}(2002)\citenamefont {Gisin},
  \citenamefont {Ribordy}, \citenamefont {Tittel},\ and\ \citenamefont
  {Zbinden}}]{QC:Gisin:2002}%
  \BibitemOpen
  \bibfield  {author} {\bibinfo {author} {\bibfnamefont {N.}~\bibnamefont
  {Gisin}}, \bibinfo {author} {\bibfnamefont {G.}~\bibnamefont {Ribordy}},
  \bibinfo {author} {\bibfnamefont {W.}~\bibnamefont {Tittel}}, \ and\ \bibinfo
  {author} {\bibfnamefont {H.}~\bibnamefont {Zbinden}},\ }\href@noop {}
  {\bibfield  {journal} {\bibinfo  {journal} {Reviews of Modern Physics}\
  }\textbf {\bibinfo {volume} {74}},\ \bibinfo {pages} {145} (\bibinfo {year}
  {2002})}\BibitemShut {NoStop}%
\bibitem [{\citenamefont {Peev}\ \emph {et~al.}(2009)\citenamefont {Peev},
  \citenamefont {Pacher}, \citenamefont {All{\'e}aume}, \citenamefont
  {Barreiro}, \citenamefont {Bouda}, \citenamefont {Boxleitner}, \citenamefont
  {Debuisschert}, \citenamefont {Diamanti}, \citenamefont {Dianati},
  \citenamefont {Dynes} \emph {et~al.}}]{peev2009secoqc}%
  \BibitemOpen
  \bibfield  {author} {\bibinfo {author} {\bibfnamefont {M.}~\bibnamefont
  {Peev}}, \bibinfo {author} {\bibfnamefont {C.}~\bibnamefont {Pacher}},
  \bibinfo {author} {\bibfnamefont {R.}~\bibnamefont {All{\'e}aume}}, \bibinfo
  {author} {\bibfnamefont {C.}~\bibnamefont {Barreiro}}, \bibinfo {author}
  {\bibfnamefont {J.}~\bibnamefont {Bouda}}, \bibinfo {author} {\bibfnamefont
  {W.}~\bibnamefont {Boxleitner}}, \bibinfo {author} {\bibfnamefont
  {T.}~\bibnamefont {Debuisschert}}, \bibinfo {author} {\bibfnamefont
  {E.}~\bibnamefont {Diamanti}}, \bibinfo {author} {\bibfnamefont
  {M.}~\bibnamefont {Dianati}}, \bibinfo {author} {\bibfnamefont
  {J.}~\bibnamefont {Dynes}},  \emph {et~al.},\ }\href@noop {} {\bibfield
  {journal} {\bibinfo  {journal} {New Journal of Physics}\ }\textbf {\bibinfo
  {volume} {11}},\ \bibinfo {pages} {075001} (\bibinfo {year}
  {2009})}\BibitemShut {NoStop}%
\bibitem [{\citenamefont {Sasaki}\ \emph {et~al.}(2011)\citenamefont {Sasaki},
  \citenamefont {Fujiwara}, \citenamefont {Ishizuka}, \citenamefont {Klaus},
  \citenamefont {Wakui}, \citenamefont {Takeoka}, \citenamefont {Miki},
  \citenamefont {Yamashita}, \citenamefont {Wang}, \citenamefont {Tanaka} \emph
  {et~al.}}]{sasaki2011field}%
  \BibitemOpen
  \bibfield  {author} {\bibinfo {author} {\bibfnamefont {M.}~\bibnamefont
  {Sasaki}}, \bibinfo {author} {\bibfnamefont {M.}~\bibnamefont {Fujiwara}},
  \bibinfo {author} {\bibfnamefont {H.}~\bibnamefont {Ishizuka}}, \bibinfo
  {author} {\bibfnamefont {W.}~\bibnamefont {Klaus}}, \bibinfo {author}
  {\bibfnamefont {K.}~\bibnamefont {Wakui}}, \bibinfo {author} {\bibfnamefont
  {M.}~\bibnamefont {Takeoka}}, \bibinfo {author} {\bibfnamefont
  {S.}~\bibnamefont {Miki}}, \bibinfo {author} {\bibfnamefont {T.}~\bibnamefont
  {Yamashita}}, \bibinfo {author} {\bibfnamefont {Z.}~\bibnamefont {Wang}},
  \bibinfo {author} {\bibfnamefont {A.}~\bibnamefont {Tanaka}},  \emph
  {et~al.},\ }\href@noop {} {\bibfield  {journal} {\bibinfo  {journal} {Optics
  Express}\ }\textbf {\bibinfo {volume} {19}},\ \bibinfo {pages} {10387}
  (\bibinfo {year} {2011})}\BibitemShut {NoStop}%
\bibitem [{\citenamefont {Dixon}\ \emph {et~al.}(2008)\citenamefont {Dixon},
  \citenamefont {Yuan}, \citenamefont {Dynes}, \citenamefont {Sharpe},\ and\
  \citenamefont {Shields}}]{dixon2008gigahertz}%
  \BibitemOpen
  \bibfield  {author} {\bibinfo {author} {\bibfnamefont {A.}~\bibnamefont
  {Dixon}}, \bibinfo {author} {\bibfnamefont {Z.}~\bibnamefont {Yuan}},
  \bibinfo {author} {\bibfnamefont {J.}~\bibnamefont {Dynes}}, \bibinfo
  {author} {\bibfnamefont {A.}~\bibnamefont {Sharpe}}, \ and\ \bibinfo {author}
  {\bibfnamefont {A.}~\bibnamefont {Shields}},\ }\href@noop {} {\bibfield
  {journal} {\bibinfo  {journal} {Optics Express}\ }\textbf {\bibinfo {volume}
  {16}},\ \bibinfo {pages} {18790} (\bibinfo {year} {2008})}\BibitemShut
  {NoStop}%
\bibitem [{\citenamefont {Liu}\ \emph {et~al.}(2010)\citenamefont {Liu},
  \citenamefont {Chen}, \citenamefont {Wang}, \citenamefont {Cai},
  \citenamefont {Wan}, \citenamefont {Chen}, \citenamefont {Wang},
  \citenamefont {Liu}, \citenamefont {Liang}, \citenamefont {Yang} \emph
  {et~al.}}]{liu2010decoy}%
  \BibitemOpen
  \bibfield  {author} {\bibinfo {author} {\bibfnamefont {Y.}~\bibnamefont
  {Liu}}, \bibinfo {author} {\bibfnamefont {T.-Y.}\ \bibnamefont {Chen}},
  \bibinfo {author} {\bibfnamefont {J.}~\bibnamefont {Wang}}, \bibinfo {author}
  {\bibfnamefont {W.-Q.}\ \bibnamefont {Cai}}, \bibinfo {author} {\bibfnamefont
  {X.}~\bibnamefont {Wan}}, \bibinfo {author} {\bibfnamefont {L.-K.}\
  \bibnamefont {Chen}}, \bibinfo {author} {\bibfnamefont {J.-H.}\ \bibnamefont
  {Wang}}, \bibinfo {author} {\bibfnamefont {S.-B.}\ \bibnamefont {Liu}},
  \bibinfo {author} {\bibfnamefont {H.}~\bibnamefont {Liang}}, \bibinfo
  {author} {\bibfnamefont {L.}~\bibnamefont {Yang}},  \emph {et~al.},\
  }\href@noop {} {\bibfield  {journal} {\bibinfo  {journal} {Optics Express}\
  }\textbf {\bibinfo {volume} {18}},\ \bibinfo {pages} {8587} (\bibinfo {year}
  {2010})}\BibitemShut {NoStop}%
\bibitem [{\citenamefont {Ursin}\ \emph {et~al.}(2007)\citenamefont {Ursin},
  \citenamefont {Tiefenbacher}, \citenamefont {Schmitt-Manderbach},
  \citenamefont {Weier}, \citenamefont {Scheidl}, \citenamefont {Lindenthal},
  \citenamefont {Blauensteiner}, \citenamefont {Jennewein}, \citenamefont
  {Perdigues}, \citenamefont {Trojek}, \citenamefont {\"{O}mer} \emph
  {et~al.}}]{Ursin:144QKD}%
  \BibitemOpen
  \bibfield  {author} {\bibinfo {author} {\bibfnamefont {R.}~\bibnamefont
  {Ursin}}, \bibinfo {author} {\bibfnamefont {F.}~\bibnamefont {Tiefenbacher}},
  \bibinfo {author} {\bibfnamefont {T.}~\bibnamefont {Schmitt-Manderbach}},
  \bibinfo {author} {\bibfnamefont {H.}~\bibnamefont {Weier}}, \bibinfo
  {author} {\bibfnamefont {T.}~\bibnamefont {Scheidl}}, \bibinfo {author}
  {\bibfnamefont {M.}~\bibnamefont {Lindenthal}}, \bibinfo {author}
  {\bibfnamefont {B.}~\bibnamefont {Blauensteiner}}, \bibinfo {author}
  {\bibfnamefont {T.}~\bibnamefont {Jennewein}}, \bibinfo {author}
  {\bibfnamefont {J.}~\bibnamefont {Perdigues}}, \bibinfo {author}
  {\bibfnamefont {P.}~\bibnamefont {Trojek}}, \bibinfo {author} {\bibfnamefont
  {B.}~\bibnamefont {\"{O}mer}},  \emph {et~al.},\ }\href@noop {} {\bibfield
  {journal} {\bibinfo  {journal} {Nature Physics}\ }\textbf {\bibinfo {volume}
  {3}},\ \bibinfo {pages} {481} (\bibinfo {year} {2007})}\BibitemShut {NoStop}%
\bibitem [{\citenamefont {Rosenberg}\ \emph {et~al.}(2009)\citenamefont
  {Rosenberg}, \citenamefont {Peterson}, \citenamefont {Harrington},
  \citenamefont {Rice}, \citenamefont {Dallmann}, \citenamefont {Tyagi},
  \citenamefont {McCabe}, \citenamefont {Nam}, \citenamefont {Baek},
  \citenamefont {Hadfield} \emph {et~al.}}]{rosenberg2009practical}%
  \BibitemOpen
  \bibfield  {author} {\bibinfo {author} {\bibfnamefont {D.}~\bibnamefont
  {Rosenberg}}, \bibinfo {author} {\bibfnamefont {C.}~\bibnamefont {Peterson}},
  \bibinfo {author} {\bibfnamefont {J.}~\bibnamefont {Harrington}}, \bibinfo
  {author} {\bibfnamefont {P.}~\bibnamefont {Rice}}, \bibinfo {author}
  {\bibfnamefont {N.}~\bibnamefont {Dallmann}}, \bibinfo {author}
  {\bibfnamefont {K.}~\bibnamefont {Tyagi}}, \bibinfo {author} {\bibfnamefont
  {K.}~\bibnamefont {McCabe}}, \bibinfo {author} {\bibfnamefont
  {S.}~\bibnamefont {Nam}}, \bibinfo {author} {\bibfnamefont {B.}~\bibnamefont
  {Baek}}, \bibinfo {author} {\bibfnamefont {R.}~\bibnamefont {Hadfield}},
  \emph {et~al.},\ }\href@noop {} {\bibfield  {journal} {\bibinfo  {journal}
  {New Journal of Physics}\ }\textbf {\bibinfo {volume} {11}},\ \bibinfo
  {pages} {045009} (\bibinfo {year} {2009})}\BibitemShut {NoStop}%
\bibitem [{\citenamefont {Renner}(2005)}]{renner2005security}%
  \BibitemOpen
  \bibfield  {author} {\bibinfo {author} {\bibfnamefont {R.}~\bibnamefont
  {Renner}},\ }\href@noop {} {\bibfield  {journal} {\bibinfo  {journal} {PhD
  Thesis, ETH No.16242, arXiv: quant-ph/0512258}\ } (\bibinfo {year}
  {2005})}\BibitemShut {NoStop}%
\bibitem [{\citenamefont {Cai}\ and\ \citenamefont
  {Scarani}(2009)}]{cai2009finite}%
  \BibitemOpen
  \bibfield  {author} {\bibinfo {author} {\bibfnamefont {R.~Y.}\ \bibnamefont
  {Cai}}\ and\ \bibinfo {author} {\bibfnamefont {V.}~\bibnamefont {Scarani}},\
  }\href@noop {} {\bibfield  {journal} {\bibinfo  {journal} {New Journal of
  Physics}\ }\textbf {\bibinfo {volume} {11}},\ \bibinfo {pages} {045024}
  (\bibinfo {year} {2009})}\BibitemShut {NoStop}%
\bibitem [{\citenamefont {Ma}\ \emph {et~al.}(2007)\citenamefont {Ma},
  \citenamefont {Fung},\ and\ \citenamefont {Lo}}]{ma2007quantum}%
  \BibitemOpen
  \bibfield  {author} {\bibinfo {author} {\bibfnamefont {X.}~\bibnamefont
  {Ma}}, \bibinfo {author} {\bibfnamefont {C.-H.~F.}\ \bibnamefont {Fung}}, \
  and\ \bibinfo {author} {\bibfnamefont {H.-K.}\ \bibnamefont {Lo}},\
  }\href@noop {} {\bibfield  {journal} {\bibinfo  {journal} {Physical Review
  A}\ }\textbf {\bibinfo {volume} {76}},\ \bibinfo {pages} {012307} (\bibinfo
  {year} {2007})}\BibitemShut {NoStop}%
\bibitem [{\citenamefont {Mayers}(2001)}]{Mayers:2001}%
  \BibitemOpen
  \bibfield  {author} {\bibinfo {author} {\bibfnamefont {D.}~\bibnamefont
  {Mayers}},\ }\href@noop {} {\bibfield  {journal} {\bibinfo  {journal}
  {Journal of the ACM (JACM)}\ }\textbf {\bibinfo {volume} {48}},\ \bibinfo
  {pages} {351} (\bibinfo {year} {2001})}\BibitemShut {NoStop}%
\bibitem [{\citenamefont {Lo}\ and\ \citenamefont {Chau}(1999)}]{Lo:1999}%
  \BibitemOpen
  \bibfield  {author} {\bibinfo {author} {\bibfnamefont {H.-K.}\ \bibnamefont
  {Lo}}\ and\ \bibinfo {author} {\bibfnamefont {H.}~\bibnamefont {Chau}},\
  }\href@noop {} {\bibfield  {journal} {\bibinfo  {journal} {Science}\ }\textbf
  {\bibinfo {volume} {283}},\ \bibinfo {pages} {2050} (\bibinfo {year}
  {1999})}\BibitemShut {NoStop}%
\bibitem [{\citenamefont {Shor}\ and\ \citenamefont
  {Preskill}(2000)}]{Shor:2000}%
  \BibitemOpen
  \bibfield  {author} {\bibinfo {author} {\bibfnamefont {P.}~\bibnamefont
  {Shor}}\ and\ \bibinfo {author} {\bibfnamefont {J.}~\bibnamefont
  {Preskill}},\ }\href@noop {} {\bibfield  {journal} {\bibinfo  {journal}
  {Physical Review Letters}\ }\textbf {\bibinfo {volume} {85}},\ \bibinfo
  {pages} {441} (\bibinfo {year} {2000})}\BibitemShut {NoStop}%
\bibitem [{\citenamefont {Scarani}\ \emph {et~al.}(2009)\citenamefont
  {Scarani}, \citenamefont {Bechmann-Pasquinucci}, \citenamefont {Cerf},
  \citenamefont {Du{\v{s}}ek}, \citenamefont {L{\"u}tkenhaus},\ and\
  \citenamefont {Peev}}]{QKD:Scarani:2009}%
  \BibitemOpen
  \bibfield  {author} {\bibinfo {author} {\bibfnamefont {V.}~\bibnamefont
  {Scarani}}, \bibinfo {author} {\bibfnamefont {H.}~\bibnamefont
  {Bechmann-Pasquinucci}}, \bibinfo {author} {\bibfnamefont {N.}~\bibnamefont
  {Cerf}}, \bibinfo {author} {\bibfnamefont {M.}~\bibnamefont {Du{\v{s}}ek}},
  \bibinfo {author} {\bibfnamefont {N.}~\bibnamefont {L{\"u}tkenhaus}}, \ and\
  \bibinfo {author} {\bibfnamefont {M.}~\bibnamefont {Peev}},\ }\href@noop {}
  {\bibfield  {journal} {\bibinfo  {journal} {Reviews of Modern Physics}\
  }\textbf {\bibinfo {volume} {81}},\ \bibinfo {pages} {1301} (\bibinfo {year}
  {2009})}\BibitemShut {NoStop}%
\bibitem [{\citenamefont {Zhao}\ \emph {et~al.}(2008)\citenamefont {Zhao},
  \citenamefont {Fung}, \citenamefont {Qi}, \citenamefont {Chen},\ and\
  \citenamefont {Lo}}]{Yi:timeshift:2008}%
  \BibitemOpen
  \bibfield  {author} {\bibinfo {author} {\bibfnamefont {Y.}~\bibnamefont
  {Zhao}}, \bibinfo {author} {\bibfnamefont {C.}~\bibnamefont {Fung}}, \bibinfo
  {author} {\bibfnamefont {B.}~\bibnamefont {Qi}}, \bibinfo {author}
  {\bibfnamefont {C.}~\bibnamefont {Chen}}, \ and\ \bibinfo {author}
  {\bibfnamefont {H.-K.}\ \bibnamefont {Lo}},\ }\href@noop {} {\bibfield
  {journal} {\bibinfo  {journal} {Physical Review A}\ }\textbf {\bibinfo
  {volume} {78}},\ \bibinfo {pages} {042333} (\bibinfo {year}
  {2008})}\BibitemShut {NoStop}%
\bibitem [{\citenamefont {Xu}\ \emph {et~al.}(2010)\citenamefont {Xu},
  \citenamefont {Qi},\ and\ \citenamefont {Lo}}]{Xu:phaseremapping:2010}%
  \BibitemOpen
  \bibfield  {author} {\bibinfo {author} {\bibfnamefont {F.}~\bibnamefont
  {Xu}}, \bibinfo {author} {\bibfnamefont {B.}~\bibnamefont {Qi}}, \ and\
  \bibinfo {author} {\bibfnamefont {H.-K.}\ \bibnamefont {Lo}},\ }\href@noop {}
  {\bibfield  {journal} {\bibinfo  {journal} {New Journal of Physics}\ }\textbf
  {\bibinfo {volume} {12}},\ \bibinfo {pages} {113026} (\bibinfo {year}
  {2010})}\BibitemShut {NoStop}%
\bibitem [{\citenamefont {Lydersen}\ \emph {et~al.}(2010)\citenamefont
  {Lydersen}, \citenamefont {Wiechers}, \citenamefont {Wittmann}, \citenamefont
  {Elser}, \citenamefont {Skaar},\ and\ \citenamefont
  {Makarov}}]{Lars:nature:2010}%
  \BibitemOpen
  \bibfield  {author} {\bibinfo {author} {\bibfnamefont {L.}~\bibnamefont
  {Lydersen}}, \bibinfo {author} {\bibfnamefont {C.}~\bibnamefont {Wiechers}},
  \bibinfo {author} {\bibfnamefont {C.}~\bibnamefont {Wittmann}}, \bibinfo
  {author} {\bibfnamefont {D.}~\bibnamefont {Elser}}, \bibinfo {author}
  {\bibfnamefont {J.}~\bibnamefont {Skaar}}, \ and\ \bibinfo {author}
  {\bibfnamefont {V.}~\bibnamefont {Makarov}},\ }\href@noop {} {\bibfield
  {journal} {\bibinfo  {journal} {Nature Photonics}\ }\textbf {\bibinfo
  {volume} {4}},\ \bibinfo {pages} {686} (\bibinfo {year} {2010})}\BibitemShut
  {NoStop}%
\bibitem [{\citenamefont {Yuan}\ \emph
  {et~al.}(2011{\natexlab{a}})\citenamefont {Yuan}, \citenamefont {Dynes},\
  and\ \citenamefont {Shields}}]{yuan2011resilience}%
  \BibitemOpen
  \bibfield  {author} {\bibinfo {author} {\bibfnamefont {Z.}~\bibnamefont
  {Yuan}}, \bibinfo {author} {\bibfnamefont {J.}~\bibnamefont {Dynes}}, \ and\
  \bibinfo {author} {\bibfnamefont {A.}~\bibnamefont {Shields}},\ }\href@noop
  {} {\bibfield  {journal} {\bibinfo  {journal} {Applied physics letters}\
  }\textbf {\bibinfo {volume} {98}},\ \bibinfo {pages} {231104} (\bibinfo
  {year} {2011}{\natexlab{a}})}\BibitemShut {NoStop}%
\bibitem [{\citenamefont {Lydersen}\ \emph {et~al.}(2011)\citenamefont
  {Lydersen}, \citenamefont {Makarov},\ and\ \citenamefont
  {Skaar}}]{lydersen2011comment}%
  \BibitemOpen
  \bibfield  {author} {\bibinfo {author} {\bibfnamefont {L.}~\bibnamefont
  {Lydersen}}, \bibinfo {author} {\bibfnamefont {V.}~\bibnamefont {Makarov}}, \
  and\ \bibinfo {author} {\bibfnamefont {J.}~\bibnamefont {Skaar}},\
  }\href@noop {} {\bibfield  {journal} {\bibinfo  {journal} {Applied physics
  letters}\ }\textbf {\bibinfo {volume} {99}},\ \bibinfo {pages} {196101}
  (\bibinfo {year} {2011})}\BibitemShut {NoStop}%
\bibitem [{\citenamefont {Yuan}\ \emph
  {et~al.}(2011{\natexlab{b}})\citenamefont {Yuan}, \citenamefont {Dynes},\
  and\ \citenamefont {Shields}}]{yuan2011reply}%
  \BibitemOpen
  \bibfield  {author} {\bibinfo {author} {\bibfnamefont {Z.}~\bibnamefont
  {Yuan}}, \bibinfo {author} {\bibfnamefont {J.}~\bibnamefont {Dynes}}, \ and\
  \bibinfo {author} {\bibfnamefont {A.}~\bibnamefont {Shields}},\ }\href@noop
  {} {\bibfield  {journal} {\bibinfo  {journal} {Applied physics letters}\
  }\textbf {\bibinfo {volume} {99}},\ \bibinfo {pages} {196102} (\bibinfo
  {year} {2011}{\natexlab{b}})}\BibitemShut {NoStop}%
\bibitem [{\citenamefont {Gerhardt}\ \emph {et~al.}(2011)\citenamefont
  {Gerhardt}, \citenamefont {Liu}, \citenamefont {Lamas-Linares}, \citenamefont
  {Skaar}, \citenamefont {Kurtsiefer},\ and\ \citenamefont
  {Makarov}}]{Gerhardt:2010}%
  \BibitemOpen
  \bibfield  {author} {\bibinfo {author} {\bibfnamefont {I.}~\bibnamefont
  {Gerhardt}}, \bibinfo {author} {\bibfnamefont {Q.}~\bibnamefont {Liu}},
  \bibinfo {author} {\bibfnamefont {A.}~\bibnamefont {Lamas-Linares}}, \bibinfo
  {author} {\bibfnamefont {J.}~\bibnamefont {Skaar}}, \bibinfo {author}
  {\bibfnamefont {C.}~\bibnamefont {Kurtsiefer}}, \ and\ \bibinfo {author}
  {\bibfnamefont {V.}~\bibnamefont {Makarov}},\ }\href@noop {} {\bibfield
  {journal} {\bibinfo  {journal} {Nature Communications}\ }\textbf {\bibinfo
  {volume} {2}},\ \bibinfo {pages} {349} (\bibinfo {year} {2011})}\BibitemShut
  {NoStop}%
\bibitem [{\citenamefont {Weier}\ \emph {et~al.}(2011)\citenamefont {Weier},
  \citenamefont {Krauss}, \citenamefont {Rau}, \citenamefont {Fuerst},
  \citenamefont {Nauerth},\ and\ \citenamefont
  {Weinfurter}}]{weier2011quantum}%
  \BibitemOpen
  \bibfield  {author} {\bibinfo {author} {\bibfnamefont {H.}~\bibnamefont
  {Weier}}, \bibinfo {author} {\bibfnamefont {H.}~\bibnamefont {Krauss}},
  \bibinfo {author} {\bibfnamefont {M.}~\bibnamefont {Rau}}, \bibinfo {author}
  {\bibfnamefont {M.}~\bibnamefont {Fuerst}}, \bibinfo {author} {\bibfnamefont
  {S.}~\bibnamefont {Nauerth}}, \ and\ \bibinfo {author} {\bibfnamefont
  {H.}~\bibnamefont {Weinfurter}},\ }\href@noop {} {\bibfield  {journal}
  {\bibinfo  {journal} {New Journal of Physics}\ }\textbf {\bibinfo {volume}
  {13}},\ \bibinfo {pages} {073024} (\bibinfo {year} {2011})}\BibitemShut
  {NoStop}%
\bibitem [{\citenamefont {Jain}\ \emph {et~al.}(2011)\citenamefont {Jain},
  \citenamefont {Wittmann}, \citenamefont {Lydersen}, \citenamefont {Wiechers},
  \citenamefont {Elser}, \citenamefont {Marquardt}, \citenamefont {Makarov},\
  and\ \citenamefont {Leuchs}}]{jain2011device}%
  \BibitemOpen
  \bibfield  {author} {\bibinfo {author} {\bibfnamefont {N.}~\bibnamefont
  {Jain}}, \bibinfo {author} {\bibfnamefont {C.}~\bibnamefont {Wittmann}},
  \bibinfo {author} {\bibfnamefont {L.}~\bibnamefont {Lydersen}}, \bibinfo
  {author} {\bibfnamefont {C.}~\bibnamefont {Wiechers}}, \bibinfo {author}
  {\bibfnamefont {D.}~\bibnamefont {Elser}}, \bibinfo {author} {\bibfnamefont
  {C.}~\bibnamefont {Marquardt}}, \bibinfo {author} {\bibfnamefont
  {V.}~\bibnamefont {Makarov}}, \ and\ \bibinfo {author} {\bibfnamefont
  {G.}~\bibnamefont {Leuchs}},\ }\href@noop {} {\bibfield  {journal} {\bibinfo
  {journal} {Physical Review Letters}\ }\textbf {\bibinfo {volume} {107}},\
  \bibinfo {pages} {110501} (\bibinfo {year} {2011})}\BibitemShut {NoStop}%
\bibitem [{\citenamefont {Li}\ \emph {et~al.}(2011)\citenamefont {Li},
  \citenamefont {Wang}, \citenamefont {Huang}, \citenamefont {Chen},
  \citenamefont {Yin}, \citenamefont {Li}, \citenamefont {Zhou}, \citenamefont
  {Liu}, \citenamefont {Zhang}, \citenamefont {Guo}, \citenamefont {Bao},\ and\
  \citenamefont {Han}}]{li2011attacking}%
  \BibitemOpen
  \bibfield  {author} {\bibinfo {author} {\bibfnamefont {H.-W.}\ \bibnamefont
  {Li}}, \bibinfo {author} {\bibfnamefont {S.}~\bibnamefont {Wang}}, \bibinfo
  {author} {\bibfnamefont {J.-Z.}\ \bibnamefont {Huang}}, \bibinfo {author}
  {\bibfnamefont {W.}~\bibnamefont {Chen}}, \bibinfo {author} {\bibfnamefont
  {Z.-Q.}\ \bibnamefont {Yin}}, \bibinfo {author} {\bibfnamefont {F.-Y.}\
  \bibnamefont {Li}}, \bibinfo {author} {\bibfnamefont {Z.}~\bibnamefont
  {Zhou}}, \bibinfo {author} {\bibfnamefont {D.}~\bibnamefont {Liu}}, \bibinfo
  {author} {\bibfnamefont {Y.}~\bibnamefont {Zhang}}, \bibinfo {author}
  {\bibfnamefont {G.-C.}\ \bibnamefont {Guo}}, \bibinfo {author} {\bibfnamefont
  {W.-S.}\ \bibnamefont {Bao}}, \ and\ \bibinfo {author} {\bibfnamefont
  {Z.-F.}\ \bibnamefont {Han}},\ }\href@noop {} {\bibfield  {journal} {\bibinfo
   {journal} {Physical Review A}\ }\textbf {\bibinfo {volume} {84}},\ \bibinfo
  {pages} {062308} (\bibinfo {year} {2011})}\BibitemShut {NoStop}%
\bibitem [{\citenamefont {Mayers}\ and\ \citenamefont
  {Yao}(2004)}]{mayers2004self}%
  \BibitemOpen
  \bibfield  {author} {\bibinfo {author} {\bibfnamefont {D.}~\bibnamefont
  {Mayers}}\ and\ \bibinfo {author} {\bibfnamefont {A.}~\bibnamefont {Yao}},\
  }\href@noop {} {\bibfield  {journal} {\bibinfo  {journal} {Quantum
  Information \& Computation}\ }\textbf {\bibinfo {volume} {4}},\ \bibinfo
  {pages} {273} (\bibinfo {year} {2004})}\BibitemShut {NoStop}%
\bibitem [{\citenamefont {Ac{\'\i}n}\ \emph {et~al.}(2007)\citenamefont
  {Ac{\'\i}n}, \citenamefont {Brunner}, \citenamefont {Gisin}, \citenamefont
  {Massar}, \citenamefont {Pironio},\ and\ \citenamefont
  {Scarani}}]{acin2007device}%
  \BibitemOpen
  \bibfield  {author} {\bibinfo {author} {\bibfnamefont {A.}~\bibnamefont
  {Ac{\'\i}n}}, \bibinfo {author} {\bibfnamefont {N.}~\bibnamefont {Brunner}},
  \bibinfo {author} {\bibfnamefont {N.}~\bibnamefont {Gisin}}, \bibinfo
  {author} {\bibfnamefont {S.}~\bibnamefont {Massar}}, \bibinfo {author}
  {\bibfnamefont {S.}~\bibnamefont {Pironio}}, \ and\ \bibinfo {author}
  {\bibfnamefont {V.}~\bibnamefont {Scarani}},\ }\href@noop {} {\bibfield
  {journal} {\bibinfo  {journal} {Physical Review Letters}\ }\textbf {\bibinfo
  {volume} {98}},\ \bibinfo {pages} {230501} (\bibinfo {year}
  {2007})}\BibitemShut {NoStop}%
\bibitem [{\citenamefont {Gisin}\ \emph {et~al.}(2010)\citenamefont {Gisin},
  \citenamefont {Pironio},\ and\ \citenamefont
  {Sangouard}}]{gisin2010proposal}%
  \BibitemOpen
  \bibfield  {author} {\bibinfo {author} {\bibfnamefont {N.}~\bibnamefont
  {Gisin}}, \bibinfo {author} {\bibfnamefont {S.}~\bibnamefont {Pironio}}, \
  and\ \bibinfo {author} {\bibfnamefont {N.}~\bibnamefont {Sangouard}},\
  }\href@noop {} {\bibfield  {journal} {\bibinfo  {journal} {Physical Review
  Letters}\ }\textbf {\bibinfo {volume} {105}},\ \bibinfo {pages} {70501}
  (\bibinfo {year} {2010})}\BibitemShut {NoStop}%
\bibitem [{\citenamefont {Lo}\ \emph {et~al.}(2012)\citenamefont {Lo},
  \citenamefont {Curty},\ and\ \citenamefont {Qi}}]{Lo:MDIQKD}%
  \BibitemOpen
  \bibfield  {author} {\bibinfo {author} {\bibfnamefont {H.-K.}\ \bibnamefont
  {Lo}}, \bibinfo {author} {\bibfnamefont {M.}~\bibnamefont {Curty}}, \ and\
  \bibinfo {author} {\bibfnamefont {B.}~\bibnamefont {Qi}},\ }\href@noop {}
  {\bibfield  {journal} {\bibinfo  {journal} {Physical Review Letters}\
  }\textbf {\bibinfo {volume} {108}},\ \bibinfo {pages} {130503} (\bibinfo
  {year} {2012})}\BibitemShut {NoStop}%
\bibitem [{\citenamefont {Tamaki}\ \emph {et~al.}(2012)\citenamefont {Tamaki},
  \citenamefont {Lo}, \citenamefont {Fung},\ and\ \citenamefont
  {Qi}}]{tamaki2012phase}%
  \BibitemOpen
  \bibfield  {author} {\bibinfo {author} {\bibfnamefont {K.}~\bibnamefont
  {Tamaki}}, \bibinfo {author} {\bibfnamefont {H.-K.}\ \bibnamefont {Lo}},
  \bibinfo {author} {\bibfnamefont {C.-H.~F.}\ \bibnamefont {Fung}}, \ and\
  \bibinfo {author} {\bibfnamefont {B.}~\bibnamefont {Qi}},\ }\href@noop {}
  {\bibfield  {journal} {\bibinfo  {journal} {Physical Review A}\ }\textbf
  {\bibinfo {volume} {85}},\ \bibinfo {pages} {042307} (\bibinfo {year}
  {2012})}\BibitemShut {NoStop}%
\bibitem [{\citenamefont {Ma}\ \emph {et~al.}(2012{\natexlab{b}})\citenamefont
  {Ma}, \citenamefont {Fung},\ and\ \citenamefont
  {Razavi}}]{ma2012statistical}%
  \BibitemOpen
  \bibfield  {author} {\bibinfo {author} {\bibfnamefont {X.}~\bibnamefont
  {Ma}}, \bibinfo {author} {\bibfnamefont {C.-H.~F.}\ \bibnamefont {Fung}}, \
  and\ \bibinfo {author} {\bibfnamefont {M.}~\bibnamefont {Razavi}},\
  }\href@noop {} {\bibfield  {journal} {\bibinfo  {journal} {Physical Review
  A}\ }\textbf {\bibinfo {volume} {86}},\ \bibinfo {pages} {052305} (\bibinfo
  {year} {2012}{\natexlab{b}})}\BibitemShut {NoStop}%
\bibitem [{\citenamefont {Wang}(2013)}]{wang2013three}%
  \BibitemOpen
  \bibfield  {author} {\bibinfo {author} {\bibfnamefont {X.-B.}\ \bibnamefont
  {Wang}},\ }\href@noop {} {\bibfield  {journal} {\bibinfo  {journal} {Physical
  Review A}\ }\textbf {\bibinfo {volume} {87}},\ \bibinfo {pages} {012320}
  (\bibinfo {year} {2013})}\BibitemShut {NoStop}%
\bibitem [{\citenamefont {Xu}\ \emph {et~al.}(2013)\citenamefont {Xu},
  \citenamefont {Curty}, \citenamefont {Qi},\ and\ \citenamefont
  {Lo}}]{Feihu:practical}%
  \BibitemOpen
  \bibfield  {author} {\bibinfo {author} {\bibfnamefont {F.}~\bibnamefont
  {Xu}}, \bibinfo {author} {\bibfnamefont {M.}~\bibnamefont {Curty}}, \bibinfo
  {author} {\bibfnamefont {B.}~\bibnamefont {Qi}}, \ and\ \bibinfo {author}
  {\bibfnamefont {H.-K.}\ \bibnamefont {Lo}},\ }\href@noop {} {\bibfield
  {journal} {\bibinfo  {journal} {arXiv:1305.6965}\ } (\bibinfo {year}
  {2013})}\BibitemShut {NoStop}%
\bibitem [{\citenamefont {Curty}\ \emph {et~al.}(2013)\citenamefont {Curty},
  \citenamefont {Xu}, \citenamefont {Cui}, \citenamefont {Lim}, \citenamefont
  {Tamaki},\ and\ \citenamefont {Lo}}]{marcos:finite:2013}%
  \BibitemOpen
  \bibfield  {author} {\bibinfo {author} {\bibfnamefont {M.}~\bibnamefont
  {Curty}}, \bibinfo {author} {\bibfnamefont {F.}~\bibnamefont {Xu}}, \bibinfo
  {author} {\bibfnamefont {W.}~\bibnamefont {Cui}}, \bibinfo {author}
  {\bibfnamefont {C.~C.~W.}\ \bibnamefont {Lim}}, \bibinfo {author}
  {\bibfnamefont {K.}~\bibnamefont {Tamaki}}, \ and\ \bibinfo {author}
  {\bibfnamefont {H.-K.}\ \bibnamefont {Lo}},\ }\href@noop {} {\bibfield
  {journal} {\bibinfo  {journal} {under preparation}\ } (\bibinfo {year}
  {2013})}\BibitemShut {NoStop}%
\bibitem [{\citenamefont {Rubenok}\ \emph {et~al.}(2013)\citenamefont
  {Rubenok}, \citenamefont {Slater}, \citenamefont {Chan}, \citenamefont
  {Lucio-Martinez},\ and\ \citenamefont {Tittel}}]{Tittel:2012:MDI:exp}%
  \BibitemOpen
  \bibfield  {author} {\bibinfo {author} {\bibfnamefont {A.}~\bibnamefont
  {Rubenok}}, \bibinfo {author} {\bibfnamefont {J.~A.}\ \bibnamefont {Slater}},
  \bibinfo {author} {\bibfnamefont {P.}~\bibnamefont {Chan}}, \bibinfo {author}
  {\bibfnamefont {I.}~\bibnamefont {Lucio-Martinez}}, \ and\ \bibinfo {author}
  {\bibfnamefont {W.}~\bibnamefont {Tittel}},\ }\href@noop {} {\bibfield
  {journal} {\bibinfo  {journal} {arXiv preprint arXiv:1304.2463}\ } (\bibinfo
  {year} {2013})}\BibitemShut {NoStop}%
\bibitem [{\citenamefont {Liu}\ \emph {et~al.}(2012)\citenamefont {Liu},
  \citenamefont {Chen}, \citenamefont {Wang}, \citenamefont {Liang},
  \citenamefont {Shentu}, \citenamefont {Wang}, \citenamefont {Cui},
  \citenamefont {Yin}, \citenamefont {Liu}, \citenamefont {Li} \emph
  {et~al.}}]{Liu:2012:MDI:exp}%
  \BibitemOpen
  \bibfield  {author} {\bibinfo {author} {\bibfnamefont {Y.}~\bibnamefont
  {Liu}}, \bibinfo {author} {\bibfnamefont {T.-Y.}\ \bibnamefont {Chen}},
  \bibinfo {author} {\bibfnamefont {L.-J.}\ \bibnamefont {Wang}}, \bibinfo
  {author} {\bibfnamefont {H.}~\bibnamefont {Liang}}, \bibinfo {author}
  {\bibfnamefont {G.-L.}\ \bibnamefont {Shentu}}, \bibinfo {author}
  {\bibfnamefont {J.}~\bibnamefont {Wang}}, \bibinfo {author} {\bibfnamefont
  {K.}~\bibnamefont {Cui}}, \bibinfo {author} {\bibfnamefont {H.-L.}\
  \bibnamefont {Yin}}, \bibinfo {author} {\bibfnamefont {N.-L.}\ \bibnamefont
  {Liu}}, \bibinfo {author} {\bibfnamefont {L.}~\bibnamefont {Li}},  \emph
  {et~al.},\ }\href@noop {} {\bibfield  {journal} {\bibinfo  {journal} {arXiv
  preprint arXiv:1209.6178}\ } (\bibinfo {year} {2012})}\BibitemShut {NoStop}%
\bibitem [{\citenamefont {Tang}\ \emph {et~al.}(2013)\citenamefont {Tang},
  \citenamefont {Liao}, \citenamefont {Xu}, \citenamefont {Qi}, \citenamefont
  {Qian},\ and\ \citenamefont {Lo}}]{zhiyuan:experiment:2013}%
  \BibitemOpen
  \bibfield  {author} {\bibinfo {author} {\bibfnamefont {Z.}~\bibnamefont
  {Tang}}, \bibinfo {author} {\bibfnamefont {Z.}~\bibnamefont {Liao}}, \bibinfo
  {author} {\bibfnamefont {F.}~\bibnamefont {Xu}}, \bibinfo {author}
  {\bibfnamefont {B.}~\bibnamefont {Qi}}, \bibinfo {author} {\bibfnamefont
  {L.}~\bibnamefont {Qian}}, \ and\ \bibinfo {author} {\bibfnamefont {H.-K.}\
  \bibnamefont {Lo}},\ }\href@noop {} {\bibfield  {journal} {\bibinfo
  {journal} {under preparation}\ } (\bibinfo {year} {2013})}\BibitemShut
  {NoStop}%
\bibitem [{\citenamefont {Hwang}(2003)}]{Hwang:2003}%
  \BibitemOpen
  \bibfield  {author} {\bibinfo {author} {\bibfnamefont {W.}~\bibnamefont
  {Hwang}},\ }\href@noop {} {\bibfield  {journal} {\bibinfo  {journal}
  {Physical Review Letters}\ }\textbf {\bibinfo {volume} {91}},\ \bibinfo
  {pages} {57901} (\bibinfo {year} {2003})}\BibitemShut {NoStop}%
\bibitem [{\citenamefont {Lo}\ \emph {et~al.}(2005{\natexlab{a}})\citenamefont
  {Lo}, \citenamefont {Ma},\ and\ \citenamefont {Chen}}]{Lo:2005}%
  \BibitemOpen
  \bibfield  {author} {\bibinfo {author} {\bibfnamefont {H.-K.}\ \bibnamefont
  {Lo}}, \bibinfo {author} {\bibfnamefont {X.}~\bibnamefont {Ma}}, \ and\
  \bibinfo {author} {\bibfnamefont {K.}~\bibnamefont {Chen}},\ }\href@noop {}
  {\bibfield  {journal} {\bibinfo  {journal} {Physical Review Letters}\
  }\textbf {\bibinfo {volume} {94}},\ \bibinfo {pages} {230504} (\bibinfo
  {year} {2005}{\natexlab{a}})}\BibitemShut {NoStop}%
\bibitem [{\citenamefont {Wang}(2005)}]{Wang:2005}%
  \BibitemOpen
  \bibfield  {author} {\bibinfo {author} {\bibfnamefont {X.}~\bibnamefont
  {Wang}},\ }\href@noop {} {\bibfield  {journal} {\bibinfo  {journal} {Physical
  Review Letters}\ }\textbf {\bibinfo {volume} {94}},\ \bibinfo {pages}
  {230503} (\bibinfo {year} {2005})}\BibitemShut {NoStop}%
\bibitem [{Note1()}]{Note1}%
  \BibitemOpen
  \bibinfo {note} {This also implies the feasibility of ``Pentagon Using China
  Satellite for U.S.-Africa Command''. See
  http://www.bloomberg.com/news/2013-04-29/pentagon-using-china-satellite-for-%
u-s-africa-command.html.}\BibitemShut {Stop}%
\bibitem [{\citenamefont {Hughes}\ \emph {et~al.}(2013)\citenamefont {Hughes},
  \citenamefont {Nordholt}, \citenamefont {McCabe}, \citenamefont {Newell},
  \citenamefont {Peterson},\ and\ \citenamefont {Somma}}]{hughes2013network}%
  \BibitemOpen
  \bibfield  {author} {\bibinfo {author} {\bibfnamefont {R.~J.}\ \bibnamefont
  {Hughes}}, \bibinfo {author} {\bibfnamefont {J.~E.}\ \bibnamefont
  {Nordholt}}, \bibinfo {author} {\bibfnamefont {K.~P.}\ \bibnamefont
  {McCabe}}, \bibinfo {author} {\bibfnamefont {R.~T.}\ \bibnamefont {Newell}},
  \bibinfo {author} {\bibfnamefont {C.~G.}\ \bibnamefont {Peterson}}, \ and\
  \bibinfo {author} {\bibfnamefont {R.~D.}\ \bibnamefont {Somma}},\ }\href@noop
  {} {\bibfield  {journal} {\bibinfo  {journal} {arXiv:1305.0305}\ } (\bibinfo
  {year} {2013})}\BibitemShut {NoStop}%
\bibitem [{\citenamefont {Lo}\ \emph {et~al.}(2005{\natexlab{b}})\citenamefont
  {Lo}, \citenamefont {Chau},\ and\ \citenamefont
  {Ardehali}}]{lo2005efficient}%
  \BibitemOpen
  \bibfield  {author} {\bibinfo {author} {\bibfnamefont {H.-K.}\ \bibnamefont
  {Lo}}, \bibinfo {author} {\bibfnamefont {H.-F.}\ \bibnamefont {Chau}}, \ and\
  \bibinfo {author} {\bibfnamefont {M.}~\bibnamefont {Ardehali}},\ }\href@noop
  {} {\bibfield  {journal} {\bibinfo  {journal} {Journal of Cryptology}\
  }\textbf {\bibinfo {volume} {18}},\ \bibinfo {pages} {133} (\bibinfo {year}
  {2005}{\natexlab{b}})}\BibitemShut {NoStop}%
\bibitem [{\citenamefont {Sangouard}\ \emph {et~al.}(2011)\citenamefont
  {Sangouard}, \citenamefont {Simon}, \citenamefont {De~Riedmatten},\ and\
  \citenamefont {Gisin}}]{sangouard2011quantum}%
  \BibitemOpen
  \bibfield  {author} {\bibinfo {author} {\bibfnamefont {N.}~\bibnamefont
  {Sangouard}}, \bibinfo {author} {\bibfnamefont {C.}~\bibnamefont {Simon}},
  \bibinfo {author} {\bibfnamefont {H.}~\bibnamefont {De~Riedmatten}}, \ and\
  \bibinfo {author} {\bibfnamefont {N.}~\bibnamefont {Gisin}},\ }\href@noop {}
  {\bibfield  {journal} {\bibinfo  {journal} {Reviews of Modern Physics}\
  }\textbf {\bibinfo {volume} {83}},\ \bibinfo {pages} {33} (\bibinfo {year}
  {2011})}\BibitemShut {NoStop}%
\bibitem [{\citenamefont {Hong}\ \emph {et~al.}(1987)\citenamefont {Hong},
  \citenamefont {Ou},\ and\ \citenamefont {Mandel}}]{HOM:1987}%
  \BibitemOpen
  \bibfield  {author} {\bibinfo {author} {\bibfnamefont {C.}~\bibnamefont
  {Hong}}, \bibinfo {author} {\bibfnamefont {Z.}~\bibnamefont {Ou}}, \ and\
  \bibinfo {author} {\bibfnamefont {L.}~\bibnamefont {Mandel}},\ }\href@noop {}
  {\bibfield  {journal} {\bibinfo  {journal} {Physical Review Letters}\
  }\textbf {\bibinfo {volume} {59}},\ \bibinfo {pages} {2044} (\bibinfo {year}
  {1987})}\BibitemShut {NoStop}%
\bibitem [{Note2()}]{Note2}%
  \BibitemOpen
  \bibinfo {note} {A threshold detector can only tell whether the input signal
  is vacuum or non-vacuum.}\BibitemShut {Stop}%
\bibitem [{Note3()}]{Note3}%
  \BibitemOpen
  \bibinfo {note} {See supplementary material at [URL will be inserted by AIP]
  for the model of the system.}\BibitemShut {Stop}%
\bibitem [{\citenamefont {Ohashi}\ \emph {et~al.}(1992)\citenamefont {Ohashi},
  \citenamefont {Shiraki},\ and\ \citenamefont {Tajima}}]{ohashi1992optical}%
  \BibitemOpen
  \bibfield  {author} {\bibinfo {author} {\bibfnamefont {M.}~\bibnamefont
  {Ohashi}}, \bibinfo {author} {\bibfnamefont {K.}~\bibnamefont {Shiraki}}, \
  and\ \bibinfo {author} {\bibfnamefont {K.}~\bibnamefont {Tajima}},\
  }\href@noop {} {\bibfield  {journal} {\bibinfo  {journal} {Lightwave
  Technology, Journal of}\ }\textbf {\bibinfo {volume} {10}},\ \bibinfo {pages}
  {539} (\bibinfo {year} {1992})}\BibitemShut {NoStop}%
\bibitem [{\citenamefont {Marsili}\ \emph {et~al.}(2013)\citenamefont
  {Marsili}, \citenamefont {Verma}, \citenamefont {Stern}, \citenamefont
  {Harrington}, \citenamefont {Lita}, \citenamefont {Gerrits}, \citenamefont
  {Vayshenker}, \citenamefont {Baek}, \citenamefont {Shaw}, \citenamefont
  {Mirin} \emph {et~al.}}]{marsili2013detecting}%
  \BibitemOpen
  \bibfield  {author} {\bibinfo {author} {\bibfnamefont {F.}~\bibnamefont
  {Marsili}}, \bibinfo {author} {\bibfnamefont {V.}~\bibnamefont {Verma}},
  \bibinfo {author} {\bibfnamefont {J.}~\bibnamefont {Stern}}, \bibinfo
  {author} {\bibfnamefont {S.}~\bibnamefont {Harrington}}, \bibinfo {author}
  {\bibfnamefont {A.}~\bibnamefont {Lita}}, \bibinfo {author} {\bibfnamefont
  {T.}~\bibnamefont {Gerrits}}, \bibinfo {author} {\bibfnamefont
  {I.}~\bibnamefont {Vayshenker}}, \bibinfo {author} {\bibfnamefont
  {B.}~\bibnamefont {Baek}}, \bibinfo {author} {\bibfnamefont {M.}~\bibnamefont
  {Shaw}}, \bibinfo {author} {\bibfnamefont {R.}~\bibnamefont {Mirin}},  \emph
  {et~al.},\ }\href@noop {} {\bibfield  {journal} {\bibinfo  {journal} {Nature
  Photonics}\ }\textbf {\bibinfo {volume} {7}},\ \bibinfo {pages} {210}
  (\bibinfo {year} {2013})}\BibitemShut {NoStop}%
\bibitem [{\citenamefont {Kok}\ and\ \citenamefont
  {Braunstein}(2001)}]{kok2001detection}%
  \BibitemOpen
  \bibfield  {author} {\bibinfo {author} {\bibfnamefont {P.}~\bibnamefont
  {Kok}}\ and\ \bibinfo {author} {\bibfnamefont {S.~L.}\ \bibnamefont
  {Braunstein}},\ }\href@noop {} {\bibfield  {journal} {\bibinfo  {journal}
  {Physical Review A}\ }\textbf {\bibinfo {volume} {63}},\ \bibinfo {pages}
  {033812} (\bibinfo {year} {2001})}\BibitemShut {NoStop}%
\bibitem [{Note4()}]{Note4}%
  \BibitemOpen
  \bibinfo {note} {One might ask `whether it is possible to increase the key
  rate by considering \protect \emph {unequal} transmissions, \protect \textit
  {i.e.}, $t_a>t_d$ and $t_b>t_e$?' Because, in such case, one can enhance the
  brightness of PDC regardless of multi-photon pairs (increasing QBER) and thus
  improve the key rate. However, we found that unequal transmissions could not
  increase the key rate too much. The key reason is that when $t_a>t_d$, the
  multi-photon pulse of WCP combined with the channel misalignment of $L_{ad}$
  will take turns to contribute significantly to the QBER. In our simulation,
  we simultaneously optimize the channel lengths (\protect \{$L_{ad}$,
  $L_{cd}$\protect \} and \protect \{$L_{be}$, $L_{ce}$\protect \}) and the
  brightness of WCP and PDC, and finally simulate the optimal key rates shown
  by the curves in the main text.}\BibitemShut {Stop}%
\bibitem [{\citenamefont {Rarity}\ \emph {et~al.}(2005)\citenamefont {Rarity},
  \citenamefont {Tapster},\ and\ \citenamefont {Loudon}}]{Rarity:2005:HOM}%
  \BibitemOpen
  \bibfield  {author} {\bibinfo {author} {\bibfnamefont {J.}~\bibnamefont
  {Rarity}}, \bibinfo {author} {\bibfnamefont {P.}~\bibnamefont {Tapster}}, \
  and\ \bibinfo {author} {\bibfnamefont {R.}~\bibnamefont {Loudon}},\
  }\href@noop {} {\bibfield  {journal} {\bibinfo  {journal} {Journal of Optics
  B: Quantum and Semiclassical Optics}\ }\textbf {\bibinfo {volume} {7}},\
  \bibinfo {pages} {S171} (\bibinfo {year} {2005})}\BibitemShut {NoStop}%
\bibitem [{Note5()}]{Note5}%
  \BibitemOpen
  \bibinfo {note} {Notice that for free-space transmission in visible
  wavelength, silicon single-photon detector can be used, which have higher
  efficiency, typically over 50\%.}\BibitemShut {Stop}%
\bibitem [{\citenamefont {Ma}\ \emph {et~al.}(2005)\citenamefont {Ma},
  \citenamefont {Qi}, \citenamefont {Zhao},\ and\ \citenamefont
  {Lo}}]{ma2005practical}%
  \BibitemOpen
  \bibfield  {author} {\bibinfo {author} {\bibfnamefont {X.}~\bibnamefont
  {Ma}}, \bibinfo {author} {\bibfnamefont {B.}~\bibnamefont {Qi}}, \bibinfo
  {author} {\bibfnamefont {Y.}~\bibnamefont {Zhao}}, \ and\ \bibinfo {author}
  {\bibfnamefont {H.-K.}\ \bibnamefont {Lo}},\ }\href@noop {} {\bibfield
  {journal} {\bibinfo  {journal} {Physical Review A}\ }\textbf {\bibinfo
  {volume} {72}},\ \bibinfo {pages} {012326} (\bibinfo {year}
  {2005})}\BibitemShut {NoStop}%
\end{thebibliography}%

\end{document}